\documentclass[12pt]{article}\usepackage{a4}
\usepackage{amssymb}
\usepackage[leftcaption]{sidecap}
\newcommand{\beao}{\begin{eqnarray*}}
\newcommand{\eeao}{\end{eqnarray*}}
\newcommand{\be}{\begin{equation}}\newcommand{\ee}{\end{equation}}
\newcommand{\bea}{\begin{eqnarray}}
\newcommand{\eea}{\end{eqnarray}}
\newcommand{\beq}{\begin{eqnarray}}
\newcommand{\eeq}{\end{eqnarray}}
\newcommand{\nn}{\nonumber}

\newcommand{\ep}{\varepsilon}
\newcommand{\ka}{\varkappa}
\newcommand{\om}{\omega}
\newcommand{\al}{\alpha}
\newcommand{\Ref}[1]{(\ref{#1})}

\usepackage{graphicx}
\usepackage{wrapfig}
\author{M. Bordag\thanks{bordag@itp.uni-leipzig.de}\\
{\small Universit\"{a}t Leipzig, Institute for Theoretical Physics}\\
{\small        Box 100 920, 04009 Leipzig, Germany }}
\title{Electromagnetic vacuum energy for two parallel slabs in terms of surface, waveguide and photonic modes}
\date{\small  January 5th, 2012}
\begin{document}
\maketitle
\begin{abstract}
The formulation of the Lifshitz formula in terms of real frequencies is reconsidered for half--spaces described by the plasma model. It is shown that besides the surface modes (for the TM polarization), and the photonic modes, also waveguide modes must be considered.
\end{abstract}
\thispagestyle{empty}
\section{Introduction}
\begin{wrapfigure}{r}{0.4\textwidth}
\centering \includegraphics[height=3.5cm,angle=270]{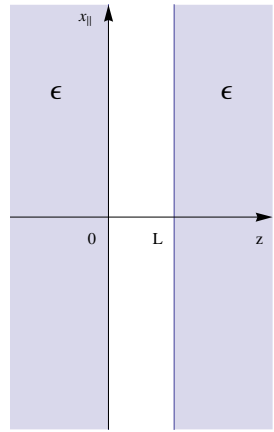}
\caption{The configuration of two half--spaces with permittivity $\ep$, separated by a gap of width $L$.}\label{fig1}
\end{wrapfigure}

In this paper the vacuum energy of the electromagnetic field in the pre\-sence of two half--spaces characterized by a permittivity $\ep$ and separated by a gap (see  Fig. \ref{fig1}), will be considered in terms of the frequencies $\om_J$ of the physical modes, symbolically
\be\label{E1}
    E_0=\frac{\hbar}{2}\sum_J\,\om_J\,.
\ee
This is the vacuum energy, defined as the half sum of the energies of all excitations, numbered by an index $J$, in the sense of zero point energy following the approach used by Casimir in \cite{casi48-51-793}. At once, for the considered configuration, this is equivalent to the well known Lifshitz formula \cite{lifs56-2-73}, which is usually written in terms of imaginary frequencies, $\om=i\xi$.

There are the following reasons to do this.

First, the corresponding mode sum (see Eq. \Ref{E2}, below) was never written down correctly for the plasma model. The point is that, besides the modes corresponding to surface waves and traveling waves (photonic modes), in \Ref{E1} also   waveguide modes must be included.

A second reason comes from a situation where  the permittivity $\ep(\om)$ is frequency dependent and  known for real frequencies $\om$ only (e.g., from experimental data). In such case, the analytic continuation to $\ep(i\xi)$ may introduce additional complications and a real frequency representation is of interest.

A third reason comes from the discussion of the contribution of surface modes to the Casimir effect a few years ago in \cite{intr05-94-110404,bord06-39-6173} which might need to be reconsidered.

The Lifshitz formula, written in terms of imaginary frequencies, has the significant advantage to involve only fast converging integrals. In contrast, integrations over the real frequency axis are notoriously difficult to handle because of the inherent oscillations. It will shown how this problem can be handled.

In the following  we consider a frequency dependent permittivity as given by the plasma model,
\be\label{ep1}  \ep(\om)=1-\frac{\om_p^2}{\om^2},
\ee
where $\om_p$ is the plasma frequency. For the considered configuration,  the regularized, but yet unrenormalized (equipped with a tilde) expression for vacuum energy, following from \Ref{E1},  reads,
\be\label{E2}   \tilde{E}_0=\frac{\hbar}{2} \int\frac{d\mathbf{k}_{||}}{(2\pi)^2}
        \left\{\sum_{sf}\om_{sf}^{1-2s}+\sum_j\om_j^{1-2s}
        +\int_{\sqrt{\om_p^2+k_{||}^2}}^\infty\frac{d\om}{\pi}\,\om^{1-2s}\,\frac{d}{d\om}\,\delta  \right\},
\ee
where $s>\frac{3}{2}$ is the regularization parameter (in fact, $\tilde{E}_0$ is the zeta function for the considered configuration). The modes include, for the TM polarization, the two surface modes with frequencies  $\om_{sf}$, and, for both polarizations,  the waveguide modes with frequencies $\om_j$ ($j=1,\dots,[\frac{\om_p}{\pi}]+1$) and the photonic modes with frequency $\om\ge\sqrt{\om_p^2+k_{||}^2 }$ (in units with $c=1$). The latter frequencies constitute, for fixed $\mathbf{k}_{||}$,  the continuous part of the spectrum and enter with a weight given by the scattering phase shift $\delta$ (its derivative is the mode density), see Eqs. \Ref{delta} and \Ref{delta1}. The vector $\mathbf{k}_{||}$ denotes the momentum in parallel to the interface. The details will be given in Section 2. It is natural that the vacuum energy needs for a renormalization. It is only the force,
\be\label{F}    F=-\frac{d}{dL}\,{E}_0,
\ee
which is finite by itself (however, only after subtraction of the empty space contribution). In Section 2 we will see  that the renormalization gives a useful suggestion  for the treatment of the sums in \Ref{E2}.

The vacuum energy, following after  renormalization from \Ref{E2}, must coincide with the known representation in terms of integration over imaginary frequencies which can be written in the form,
\be\label{LF}   E_0=\frac{1}{2} \int\frac{d\mathbf{k}_{||}}{(2\pi)^2}
    \int_0^\infty\frac{d\xi}{\pi}\, \xi \frac{d}{d\xi}
    \left(\ln \frac{1}{1-r_{\rm TE}^2e^{-2\eta L}}
          +\ln\frac{1}{1-r_{\rm TM}^2e^{-2\eta L}}        \right),
\ee
with the
reflection coefficients,
\be\label{r}     r_{\rm TE}=\frac{\ka-\eta}{    \ka+     \eta},\quad
                 r_{\rm TM}=\frac{\ka-\ep(i\xi)\eta}{\ka+\ep(i\xi)\eta},
\ee
and the notations $\ka=\sqrt{\ep(i\xi)\xi^2+k_{||}^2}$, $\eta=\sqrt{\xi^2+k_{||}^2}$. The integrations in \Ref{LF} converge  fast due to the exponential factors. Therefore no renormalization or regularization is necessary in this formula. Also, it can be seen, that it is normalized in the correct way as to be vanishing for infinite separation. In fact, for large separation $L$, one can put the reflection coefficients \Ref{r} equal to unity (formally by $\om_p\to\infty$) and will recover the result originally obtained by Casimir in \cite{casi48-51-793}.

Eq. \Ref{LF}, or the force derived from it by means of \Ref{F}, is one of the many ways of representing the {\it Lifshitz formula} in terms of imaginary frequencies (see, e.g., Eq.(12.29) in \cite{BKMM}, for reference).
Other representations of this kind differ only by changed notations, by substitutions of the integration variables, by integrating by parts in $\xi$ or by considering finite temperature $T$. The latter can be achieved by substituting the integration over $\xi$ by the Matsubara sum.

In general, the Lifshitz formula is used for more general permittivities as \Ref{ep1} is. Thereby it is always assumed that $\ep(i\xi)$   is real such that the energy \Ref{LF} is real too. There are, however, also representations of $E_0$ with integration over the real $\om$-axis, see Eq. (2.4) in \cite{lifs56-2-73} or Eq. (12.37) in \cite{BKMM}, for example. These representations look like Eq. \Ref{E2} with the $\om$-integration starting from $\om=0$ and no sums. It must be mentioned that these representations were derived under the assumption, motivated by dissipation or other physical reasons, that  ${\rm Im}(\ep(\om))>0$ holds and  that there are no poles of the transmission coefficients on the real $\om$-axis. Such systems do not have eigenvalues in terms of real frequencies $\om$ in the sense as discussed above (see also the discussion in \cite{bord11-71-1788}). For this reason, there is no direct relation between such representations and Eq. \Ref{E2}.

The Lifshitz formula is of fundamental importance in many areas of modern physics, including adhesion forces, atom-wall interactions, van der Waals forces and its  applications reach to nano--technology and biology. Accordingly, this formula has a long history. Originally it was obtained in \cite{lifs56-2-73} from the energy of the electromagnetic field driven by fluctuating charges in the half--spaces, the latter being thermally averaged following the theory by Rytov. In \cite{lifs56-2-73}, also the limiting cases were derived. For large separation, Casimir's result was shown to follow. For small separation, a new formula for the van der Waals force, acting between  half--spaces, emerged  from \Ref{LF} by putting in  \Ref{r} $\om_p\to0$ in $\ka$. The corresponding interaction energy can be written in the form
\be\label{vdW}   E_0=\frac{\hbar}{2} \int\frac{d\mathbf{k}_{||}}{(2\pi)^2}
    \int_0^\infty\frac{d\xi}{\pi}\, \xi \frac{d}{d\xi}
              \ln\frac{1}{1-\left(\frac{\ep(i\xi)-1}{\ep(i\xi)+1}\right)^2e^{-2k_{||} L}}  .
\ee
By rarefying the media in the half--spaces, also the known Casimir-Polder and the London forces between molecules emerged as limiting cases.

It must be mentioned that the understanding of the Casimir and van der Waals forces appeared in
\cite{casi48-51-793} and in  \cite{lifs56-2-73} from completely different ideas. While Casimir started from the zero--point energy of the electromagnetic field, Lifshitz considered the fluctuations in the medium as source. In the modern understanding these two are equivalent. However, the discussion about  two ways continues until present time, see  for example \cite{jaff05-72-021301}.

In the historical development, the next step after \cite{lifs56-2-73} was made in \cite{dzya61-73-381}, by considering the problem in the general framework of thermal quantum field theory. This procedure was found very difficult and lengthy. Also, there was an interest in understanding the Lifshitz formula in terms of vacuum energy in the aim of Casimir.
An important step in this direction was made by the very influential paper  \cite{vank68-26-307}
in the non--retarded case starting from the vacuum energy \Ref{E1}. For small separation, when the electromagnetic interaction happens instantaneously, the dominating contribution to the mode sum in \Ref{E2} comes from the surface modes present in the TM polarization.
From these, in \cite{vank68-26-307}, the limiting case \Ref{vdW} of the Lifshitz formula  was obtained in a  short derivation on 2 pages. Soon after, this approach was generalized to the non--retarded case. In \cite{ninh70-2-323}, the idea in \cite{vank68-26-307} of the vacuum energy as resulting from the surface modes was taken over literally to the non--retarded case by accounting for the full momentum dependence in the reflection coefficients \Ref{r}. The sum over these two modes in \Ref{E2} was transformed into an integral over imaginary frequencies and formula \Ref{LF} was reobtained. Independently, in  \cite{gerl71-4-393}, this result was obtained by the same method, admitting, however, that, besides the surface modes, all evanescent modes should be included.

It must be mentioned, that in the papers \cite{ninh70-2-323} and \cite{gerl71-4-393} the contributions from the photonic modes were missed.  This was compensated by ignoring the cut in the complex frequency plane the functions generating the dispersion relations have. These omissions were observed quite easily, for example, in \cite{lang73-58-1176} and in \cite{schr73-43-282} (see also, the thesis  \cite{schram75}\footnote{This thesis can be ordered from the library of the University of Utrecht, the  book-number is UB-ND MAG DISS UTRECHT QU 1975-25.},  p. 76), but did not receive due attention. This is another reason for the present paper to clarify in detail the role of the various modes in \Ref{E2}. It must be mentioned, that the waveguide modes, including those in planar waveguides,  are well known in electrodynamics, see \cite{marc93b}, for example. However, their role in the Casimir effect was, to the authors knowledge,  never considered. One exception is the paper \cite{intr07-76-033820}, where these modes were discussed (using different naming, see  Eq. (8) there)\footnote{I'm indebted to the authors of that paper for the hint.}.

In the following section we consider the basic formulas for determining the modes and point out their basic properties. In Section 3 we consider the vacuum energy, separately for both polarizations. The last section has some conclusions. In the appendices we give a short description of the reverse transitions, from ima\-gi\-nary to real frequencies.

\noindent In the following, we use units with $\hbar=c=1$.

\section{Basic equations and  modes}

In this section we collect the basic formulas from classical electrodynamics and identify the modes. We do this in quite big detail. Although all these questions were considered and solved long ago, there is still some confusion about the question which modes are relevant for the Casimir effect in the given configuration.

We consider two parallel half--spaces characterized by a permittivity $\ep(\om)$ (see Fig. \ref{fig1}). These are separated by an empty gap of width $L$ with parallel interfaces and we have $\ep=1$ in the gap. As it is well known, the solutions of the Maxwell equations for this configuration separate into TE and TM modes. After taking  Fourier transform in the time $t$ and in the directions $\mathbf{x}_{||}$ parallel to the interfaces, the electric field strength $\cal E$ becomes proportional to
\be\label{Fz1}  {\cal E}\sim e^{-i\om t+i \mathbf{k}_{||}\mathbf{x}_{||}}\, \Phi(z)
\ee
and the Maxwell equations reduce to the equation
\be\label{Fz2}  \left(\ep(\om,z)\,\om^2+k_{||}^2-\frac{d^2}{dz^2}\right)\Phi(z)=0,
\ee
where $\om$ is the frequency and $\mathbf{k}_{||}$ is the two--dimensional momentum in direction parallel to the interface. Since Eq. \Ref{Fz2} holds on the whole $z$-axis, we have to consider the permittivity $\ep(\om,z)$ with
$\ep(\om,z)=1$ for $z\in [0,L]$ and
$\ep(\om,z)=\ep(\om)$ for $z\notin [0,L]$.
The components of the function $\Phi(z)$ have to fulfill the well known matching conditions on the interfaces. For example,  for the components corresponding to the electric field parallel to the interface
these demand
\be\begin{array}{llrl}\label{Mc}
    \Phi(z)\quad & \mbox{and}&\quad \frac{d}{dz}\,\Phi(z) \qquad &\mbox{(TE mode)}  \\[10pt]
    \Phi(z) \quad& \mbox{and}&\quad \ep(\om,z)\frac{d}{dz}\,\Phi(z) \qquad &\mbox{(TM mode)}
\end{array}\ee
to be continuous.

The solutions of Eq. \Ref{Fz2} can be written in the form
\be\label{Fz3}  \Phi(z)=\left\{
    \begin{array}{rcc}
        e^{ikz}+R(k)\,e^{-ikz} &~~~& (z<L)   \\[8pt]
        \mu\, e^{iqz}+\nu\, e^{-iqz}    &&   (0<z<L)   \\[8pt]
        T(k)\, e^{ikz}        && (L<z)
        \end{array} \right.
\ee
and represent a wave incoming from the left. Here, $k$ has the meaning of the momentum in direction of the $z$-axis, i.e., in perpendicular to the interfaces, outside the gap, and $q$ is the corresponding momentum inside the gap. In Eq. \Ref{Fz3},  $R(k)$ and $T(k)$ are the reflection and transmission coefficients, $\mu$ and $\nu$ are some more coefficients. We indicate the momentum dependence of $R(k)$ and $T(k)$ keeping in mind that for the TM case there is an additional dependence on $k_{||}$.

From Eq. \Ref{Fz2}, i.e., from the Maxwell equations, for these momenta the relations
\bea\label{D1}  \ep(\om)\,\om^2&=&k_{||}^2+k^2,    \nn\\
                    \om^2&=&k_{||}^2+q^2
\eea
follow. For the permittivity given by Eq. \Ref{ep1}, these relations read
\bea\label{D2}  \om^2&=&\om_p^2+k_{||}^2+k^2,    \nn\\
                    \om^2&=&k_{||}^2+q^2.
\eea
It is clear that out of the four quantities, $\om$, $k_{||}$, $k$ and $q$, only two are independent. Below we will use different choices of the independent ones, the remaining must be expressed in terms of these using the above equations.

Using the matching conditions \Ref{Mc}, all coefficients in \Ref{Fz3} can be determined. In the following we need only the transmission coefficients, which take the well known form
\bea\label{T1}
    T^{\rm TE}(k)&=&\frac{4qk\,e^{-ikL}}{(q+k)^2\,e^{-iqL}-(q-k)^2\,e^{iqL}}\,,
\nn\\    T^{\rm TM}(k)&=&\frac{4\ep(\om)qk\,e^{-ikL}}{(\ep(\om) q+k)^2\,e^{-iqL}-(\ep(\om) q-k)^2\,e^{iqL}}\,,
\eea
for the two polarizations. For later use we rewrite these expression in the form
\bea\label{T2}
        T^{\rm TE}(k)&=&T_1^{\rm TE}(k)\frac{4qk}{(q+k)^2}\,e^{i(q-k)L},
\nn\\    T^{\rm TM}(k)&=&T_1^{\rm TM}(k)\frac{4\ep(\om)qk}{(\ep(\om)q+k)^2}\,e^{i(q-k)L},
\eea
with
\bea\label{T3}
    T_1^{\rm TE}(k)&=&\frac{1}{1-r_{\rm TE}^2\,e^{i2qL}}\,,
\nn\\    T_1^{\rm TM}(k)&=&\frac{1}{1-r_{\rm TM}^2\,e^{i2qL}}\,,
\eea
where
\bea\label{r2}  r_{\rm TE}&=&\frac{k-q}{k+q}\, ,  \nn \\
                r_{\rm TM}&=&\frac{k-\ep(\om)\,q}{k+\ep(\om)\,q}\,,
\eea
are the reflection coefficients on a single interface.

By Eq. \Ref{Fz2} and the matching conditions \Ref{Mc}, a spectral problem is set up. The spectrum consists of all real $\om$, allowed by these equations and conditions. This problem can be illustrated by establishing  a relation to a simple quantum mechanical Schr\"odinger equation by rewriting \Ref{Fz2} in the form
\be\label{qm1}  \left(\-\frac{d^2}{dz^2}+V(z)\right)\Phi(z)=\om^2 \Phi(z),
\ee
with a potential
\be\label{qm2}
        V(z)=\left\{\begin{array}{rcl}\om_p^2+k_{||}^2&\quad&(z\notin [0,L]),
               \\[8pt]                     k_{||}^2 && (z\in[0,L]).
                    \end{array}\right.
\ee
With the matching conditions \Ref{Mc} for the TE mode, this is a simple exercise with a piecewise constant potential, the so called {\it finite square well}. For the TM mode the potential is more complicated. But anyway, we can determine the spectrum. In relation to the electromagnetic problem, given by Eq. \Ref{Fz2}, it must be understood that we consider a problem with fixed but arbitrary $k_{||}$.

For the TE mode, the solution of the spectral problem is obvious. We have scattering states with $\om\ge\sqrt{\om_p^2+k_{||}^2}$ and bound states with $k_{||}\le\om \le \sqrt{\om_p^2+k_{||}^2}$, the latter having imaginary momenta,
\be\label{ka1}k\equiv i \ka=i\sqrt{\om_p^2+k_{||}^2-\om^2},
\ee
outside the box. As concerns  the TM mode, it is known that it has the same properties (except for some bound states which can be located in $0\le\om\le k_{||}$ too), although that is a bit more complicated to show.

From the scattering problem,  it is known that the transmission coefficients, $T^{\rm TE}(k)$ and $T^{\rm TM}(k)$, are meromorphic functions in the upper half of the complex $k$-plane with simple poles on the upper part of the imaginary axis and a continuous continuation to the real $k$-axis. The location of the poles is just the momenta $k=i\ka$ corresponding to bound states. For the function $\Phi(z)$, this results in an exponential decrease for $|z|\to\infty$. Finally, we mention the property
\be\label{Tk}T(-k)=T^*(k)
\ee
for real $k$ under complex conjugation. It is this relation which motivates us to indicate the dependence on the argument $k$ explicitly.

\begin{figure}[t]
\includegraphics[width=14.5cm]{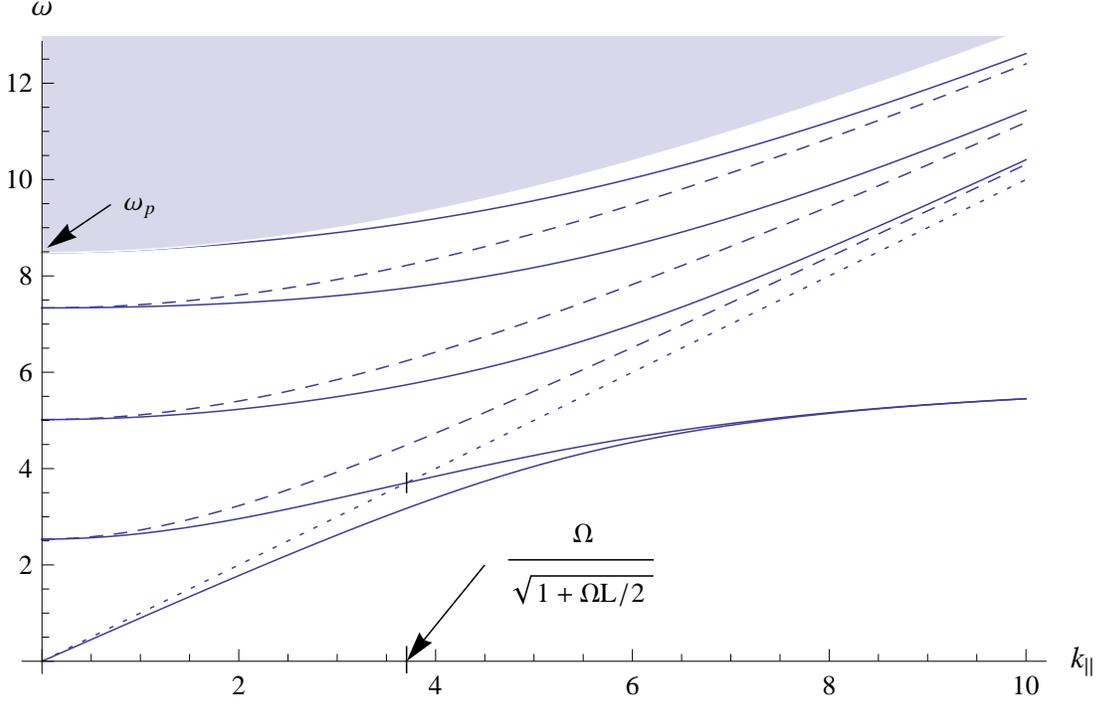}
\caption{The spectrum of the electromagnetic field for $\om_p=2.7  \pi$. Solid lines represent, from bottom to top, the frequencies of the symmetric and of the antisymmetric surface modes and, further, of the TM waveguide modes $\om_j^{\rm TM}$ ($j=1,2,3$). The dashed lines are the frequencies of the TE waveguide modes $\om_j^{\rm TE}$ ($j=1,2,3$). The dotted line corresponds to $\om=k_{||}$ and separates imaginary ($\om\le  k_{||}$) from real values of the momentum $q$ (see Eq. \Ref{D1}). The shaded region has $\om\ge\sqrt{\om_p^2+k_{||}^2}$ and represents the photonic modes.}\label{fig2}
\end{figure}

Returning to  electrodynamics, we can identify the following modes,
\begin{itemize}
    \item   photonic modes with $\om\ge\sqrt{\om_p^2+k_{||}^2}$,
    \item   surface and waveguide modes with $0\le\om\le\sqrt{\om_p^2+k_{||}^2}$.
\end{itemize}
The photonic modes correspond to the quantum mechanical scattering states and have a continuous spectrum. The waveguide and surface modes correspond to the bound states. Accounting for the momentum $k_{||}$, these form a tower of continuous  states.  It must be mentioned that all electromagnetic waves in this problem are propagating waves having real frequency $\om$. The difference between them is that the photonic modes do propagate in all spatial directions whereas the surface modes do propagate only in direction in parallel to the interface and the waveguide modes do propagate only inside the gap (including zig-zag like propagation). It should be mentioned that commonly waves with $\om<k_{||}$ are called {\it evanescent waves} (see \cite{BKMM}, p.289). In Fig. \ref{fig2} this corresponds to the region below the dotted line.


Now we consider the waveguide modes in more detail. First, we take the TE case, which is easier since the transmission coefficient $T^{\rm TE}(k)$ does not depend on $k_{||}$. As already mentioned, these modes correspond to the quantum mechanical bound states and their location is given by the poles of $T^{\rm TE}(k)$. These appear for imaginary $k=i\sqrt{\om_p^2-q^2}$. From Eqs. \Ref{T1}, or \Ref{T3} together with \Ref{r2}, the conditions
\be\label{Fc1}  \frac{\sqrt{\om_p^2-q^2}}{q}=
            \left\{\begin{array}{rl}\tan\frac{qL}{2} \quad & \mbox{(symmetric),}\\[8pt]
                                    -\cot\frac{qL}{2}\quad & \mbox{(antisymmetric),}
            \end{array}\right.
\ee
follow, which are for convenience written in terms of the variable $q$, see Eq. \Ref{D2}. The symmetry properties refer to the function $\Phi(z)$, \Ref{Fz3}, with respect to reflection on the middle of the gap. These conditions are exactly the same  as that defining the mentioned bound states. The solutions of Eqs. \Ref{Fc1} will be denoted by $q_j^{\rm TE}$ with $j=1,2,\dots,[\frac{\om_p}{\pi}]+1$, where $[\dots]$ denotes the integer part. Using \Ref{D2}, we define the corresponding frequencies,
\be\label{omj1}\om_j^{\rm TE}=\sqrt{k_{||}^2+({q_j^{\rm TE}})^2},
\ee
for which
\be\label{omj2} k_{||}\le \om_j^{\rm TE} \le \sqrt{\om_p^2+k_{||}^2}
\ee
holds. The momenta $k$, related to these solutions, are imaginary, $k\to i\ka_j^{\rm TE}$, with
\be\label{kaj1} \ka_j^{\rm TE}=\sqrt{\om_p^2-({q_j^{\rm TE}})^2}.
\ee
The numbering is done in a way that
\be\label{omj3}{\om_j^{\rm TE}}<{\om_{j+1}^{\rm TE}}
\ee
holds.
Solving Eqs. \Ref{Fc1} numerically, pictures for the frequencies $\om_j^{\rm TE}$ as function of $k_{||}$ can be generated. This is shown in Fig. \ref{fig2}. It must be mentioned that there is at least one such state for any fixed value of $\om_p$.

A characteristic property of the waveguide modes is that these, for $\om_p\to\infty$, turn into the modes known for ideally conducting walls, $q_j^{\rm TE}\to \frac{2\pi j}{L}$, with  $j=1,2,\dots$. This can be seen directly in Eq. \Ref{Fc1} and this is equivalent to an infinite square well in place of \Ref{qm2}.

Now we consider the TM modes. The condition for $T^{\rm TM}(k)$, Eq. \Ref{T1}, to have poles,
\be\label{Fc2}  \frac{\sqrt{\om_p^2-q^2}}{\ep(\om)q}=
            \left\{\begin{array}{rl}\tan\frac{qL}{2} \quad & \mbox{(symmetric)}\\[8pt]
                                    -\cot\frac{qL}{2}\quad & \mbox{(antisymmetric)},
            \end{array}\right.
\ee
differs from \Ref{Fc1} by the presence of the permittivity,
\be\label{ep2}\ep(\om)=\frac{q^2+k_{||}^2-\om_p^2}{q^2+k_{||}^2},
\ee
only, where \Ref{D2} was used to express $\om$ in terms of $q$. We denote the solutions of \Ref{Fc2} by $q_j^{\rm TM}$. For $k_{||}\ge\om_p$ we see immediately that, because of $\ep(\om)<1$ in the left hand side of \Ref{Fc2}, the TM solutions are larger than the corresponding TE solutions,
\be\label{qj1}q_j^{\rm TM}\ge q_j^{\rm TE},\qquad(j=1,2,\dots,\left[\frac{\om_p}{\pi}\right]+1).
\ee
By means of \Ref{D2}, a similar relation holds for the frequencies,
\be\label{qem}\om_j^{\rm TM}\ge\om_j^{\rm TE},
\ee
where we defined
\be\label{omj4}\om_j^{\rm TM}=\sqrt{k_{||}^2+(q_j^{\rm TM})^2}.
\ee
For $k_{||}\to\infty$, we have $\ep(\om)\to1$  and both kinds of solutions coincide. In this way, we observe a one-to-one correspondence between the TE and the TM modes. For $k_{||}<\om_p$, the picture changes since $\ep(\om)$ as a function of $q$ goes through zero. In this case the mentioned correspondence persist, but an additional solution, $q_0^{\rm TM}$ with a corresponding frequency $\om_0^{\rm TM}$ (also obeying the inequality \Ref{omj3}), appears provided $k_{||}\le\om_p/\sqrt{1+\om_p L/2}$ holds.

Now, since all $q_j^{\rm TM}$ are real, for the frequencies the inequalities
\be\label{omj5} k_{||}\le\om_j^{\rm TM}\le\sqrt{\om_p^2+k_{||}^2}
\ee
hold like Eq. \Ref{omj2} in the TE case. Also for these modes, the momenta $k\to i\ka_j^{\rm TM}$, with
\be\label{ka2}\ka_j^{\rm TM}=\sqrt{\om_p^2-(q_j^{\rm TM})^2},
\ee
are imaginary.

There is one more relation between the TE and the TM modes. For $k_{||}\to0$, Eq. \Ref{Fc2} for the TM modes turns into Eq. \Ref{Fc1} for the TE modes. Thereby the correspondence changes by one. In this way, the inequality
\be\label{corr}
\om_{j-1}^{\rm TM}\le \om_{j}^{\rm TE}\le \om_{j}^{\rm TM},\qquad (j=1,2,\dots,\left[\frac{\om_p}{\pi}\right]+1)
\ee
can be established. The left equality is reached for $k_{||}\to0$ and the right one for  $k_{||}\to\infty$.

In addition to the TM modes considered above, there are also poles of the reflection coefficient for imaginary momentum, $q\to i\eta$, for which Eqs. \Ref{Fc2} can be rewritten in the form
\be\label{Fc3}  -\frac{\sqrt{\om_p^2+\eta^2}}{\ep(\om)\eta}=
            \left\{\begin{array}{rl}\tanh\frac{qL}{2} \quad & \mbox{(symmetric),}\\[8pt]
                                    \coth\frac{qL}{2}\quad & \mbox{(antisymmetric)},
            \end{array}\right.
\ee
and the permittivity expressed in terms of $\eta$ reads
\be\label{ep3} \ep(\om)=\frac{k_{||}^2-\om_p^2-\eta^2}{k_{||}^2-\eta^2}.
\ee
Each of these equations has one solution which we denote by $\eta_{sf}$ ($sf=s,a$ according to   symmetry). The corresponding frequencies are
\be\label{omsf} \om_{sf}=\sqrt{k_{||}^2-\eta_{sf}^2}\,.
\ee
Of course, also in this case the momenta $k$ are imaginary, $k\to i \ka$, with
\be\label{kasf}\ka_{sf}=\sqrt{\om_p^2+\eta_{sf}^2}.
\ee
These two solutions, $\om_{s}$ and $\om_{a}$, correspond to waves decreasing exponentially to each side of each interface in opposite to the waveguide modes which decrease only outside the gap but oscillate inside.

It is known that, for $L\to\infty$, these solutions turn into the surface plasmons well known in the physics of metals. In this limit, the right hand side of Eq. \Ref{Fc3} equals unity and the equation allows for an explicit solution,
\be\label{omsi} \om_{single}=\sqrt{\frac{\om_p^2}{2}+k_{||}^2-\sqrt{\left(\frac{\om_p^2}{2}\right)^2+k_{||}^4}}\,.
\ee
At large separation, such solution  exists on each interface. At finite separation, the frequency $\om_{single}$ splits into two, $\om_s$ and $\om_a$, for which the inequality
\be\label{omsa}\om_s\le \om_{single}\le \om_a
\ee
holds.

There is one peculiarity about the antisymmetric surface plasmon. It is a solution of Eq. \Ref{Fc3} only if the inequality
\be\label{kpo}\frac{\om_p}{\sqrt{1+\frac{\om_p L}{2}}}\le k_{||}
\ee
holds.  In that case the momentum in the gap, $q=i\eta_a$, is imaginary. For smaller $k_{||}$, this mode matches  the solution $\om_0^{\rm TM}$ which was found among the waveguide modes if just the opposite to \Ref{kpo} holds. It is clear that these constitute one single mode which must be identified with the antisymmetric  surface mode. We mention that the question of whether to include the  lower part of the antisymmetric surface plasmon, where its frequency is given by $\om_o^{\rm TM}$, into the vacuum energy, was discussed in \cite{lena06-96-218901} and \cite{intr06-96-218902}.

The frequencies of the surface modes are shown in Fig. \ref{fig2}, too. The antisymmetric one crosses the line $\om=k_{||}$ just for $k_{||}=\om_p/\sqrt{1+\om_p L/2}$. For lower $k_{||}$, is is given by $\om_0^{\rm TM}$, Eq. \Ref{omj4},  and, for larger $k_{||}$, by $\om_a$, Eq. \Ref{omsf}.

Finally we mention, as a quite special property of the surface modes, that their frequencies have an upper bound,
\be\label{sfb}\om_{sf}\le\frac{\om_p}{\sqrt{2}}\,.
\ee
This can be seen by mentioning that $\om_{sf}$ are monoton functions increasing with increasing $k_{||}$. For large $k_{||}$, the solutions $\eta$ of the equations \Ref{Fc3} grow proportional to $k_{||}$. Consequently, the hyperbolic functions in the right hand side of these equations turn into unity and we are left with the equation for a single surface plasmon. From the corresponding energy, Eq. \Ref{omsi}, one can then infer the bound \Ref{sfb}.

\section{Vacuum energy as mode sum}
In this section we consider the vacuum energy of the electromagnetic field in the sense of Eq. \Ref{E1} and will give Eq. \Ref{E2} a precise meaning. We start with mentioning that the spectrum, if accounting for $k_{||}$, is completely continuous. Hence one needs to separate the empty space contribution. This can be done in numerous different approaches. We follow that used in \cite{bord95-28-755}. There, a large box was introduced, whose volume was subsequently tended to infinity. As a result, Eq. \Ref{E2} appears with the scattering phase shift expressed in terms of the transmission coefficient,
\be\label{delta}  \delta=\frac{1}{2i}\ln\frac{T(k)}{T(k)^*}\,.
\ee
Another, equivalent, form of writing is
\be\label{delta1}    \delta={\rm Im}(  \ln T(k)),
\ee
which is also frequently used in literature.

Because of the separation of polarizations in the considered problem, the energy consist of two parts,
\be\label{EtTETM}\tilde{E}_0=\tilde{E}^{\rm TE}+\tilde{E}^{\rm TM}.
\ee
In the next subsection we consider the contribution from the TE polarization. It is easier to handle. In a subsequent subsection,  we consider the TM case.

\subsection{TE case}
For the TE polarization we have to consider waveguide and photonic modes. Accordingly, we divide the regularized vacuum energy,
\be\label{Et1} \tilde{E}^{\rm TE}=\tilde{E}^{\rm TE}_{\rm wg}+\tilde{E}^{\rm TE}_{\rm cont},
\ee
into the waveguide contribution,
\be\label{Etwg1} \tilde{E}^{\rm TE}_{\rm wg}=
            \frac{1}{2} \int\frac{d\mathbf{k}_{||}}{(2\pi)^2}
            \sum_j\left(k_{||}^2+(q_j^{\rm TE})^2\right)^{\frac12-s},
\ee
and the contribution from the photonic modes,
\be\label{Etcont1}\tilde{E}^{\rm TE}_{\rm cont}=
             \frac{1}{2} \int\frac{d\mathbf{k}_{||}}{(2\pi)^2}
             \int_0^\infty\frac{dk}{2i\pi}\left(\om_p^2+k_{||}^2+k^2\right)^{\frac12-s}
             \frac{d}{dk}\ln\frac{T^{\rm TE}(k)}{T^{\rm TE}(k)^*},
\ee
As compared to Eq.\Ref{E2}, we changed the variable of integration from $\om$ to $k$ using \Ref{D2}. It must be mentioned that both above expressions carry ultraviolet divergencies and one needs to keep the regularization with $s>\frac32$.

Now we carry out the integration over $\mathbf{k}_{||}$, which can be done in the TE case since neither $T^{\rm TE}(k)$ nor $q_j^{\rm TE}$ depend on $\mathbf{k}_{||}$ (note $q=\sqrt{\om_p^2+k^2}$ from Eq. \Ref{D2}). The integration is simple and we come to
\be\label{Etwg2} \tilde{E}^{\rm TE}_{\rm wg}=
            \frac{-1}{4\pi(3-2s)}
            \sum_j\left(q_j^{\rm TE}\right)^{3-2s},
\ee
and
\be\label{Etcont2}  \tilde{E}^{\rm TE}_{\rm cont}=
                    \frac{-1}{4\pi(3-2s)}
             \int_0^\infty\frac{dk}{2i\pi}\left(\om_p^2+k^2\right)^{\frac32-s}
             \frac{d}{dk} \ln\frac{T^{\rm TE}(k)}{T^{\rm TE}(k)^*}\,.
\ee
In the waveguide contribution $\tilde{E}^{\rm TE}_{\rm wg}$, the analytic continuation can now be carried out simply by putting $s=0$. However, we postpone this. The second contribution will be split, using the factorization in \Ref{T2},
\be\label{Etcont3}  \tilde{E}^{\rm TE}_{\rm cont}= {E}^{\rm TE}_{\rm cont}+E_{L_1},
\ee
into
\be\label{Econt}  {E}^{\rm TE}_{\rm cont}=
                    \frac{-1}{4\pi(3-2s)}
             \int_0^\infty\frac{dk}{2i\pi}\left(\om_p^2+k^2\right)^{\frac32-s}
             \frac{d}{dk}\ln\frac{T^{\rm TE}_1(k)}{T^{\rm TE}_1(k)^*}\,,
\ee
and
\be\label{EL1_1}  {E}_{L_1}=
                    \frac{-L}{42\pi(3-2s)}
             \int_0^\infty\frac{dk}{\pi}\left(\om_p^2+k^2\right)^{\frac32-s}
             \frac{d}{dk} \,(q-k)\,.
\ee
Note that the factor $4qk/(k+q)^2$ dropped out since $k$ and $q$ are real here.

The latter contribution is proportional to the width $L$ of the gap and it can be calculated explicitly,
\be\label{EL1_2} {E}_{L_1}=-\frac{\om_p^{2(2-s)}L}{12\pi^2}\,h_1(s),
\ee
with
\be\label{h1}   h_1(s)=\frac{1}{2(s-2)}-\frac{\sqrt{\pi}\,\Gamma(s-2)}{2\Gamma(s-\frac32)}.
\ee
The function $h_1(s)$ has a pole in $s=0$,
\be\label{h1p}  h_1(s)=\frac{-3}{16s}+O(1),
\ee
which results from ultraviolet divergence.

In the first contribution, $ {E}^{\rm TE}_{\rm cont}$, it is possible to put $s=0$,
\be\label{Econt1}  {E}^{\rm TE}_{\rm cont}=
                    \frac{-1}{12\pi}
             \int_0^\infty\frac{dk}{\pi}\left(\om_p^2+k^2\right)^{\frac32}
             \frac{d}{dk}\, {\rm Im}(\ln{T_1(k)})\,,
\ee
since the integral is convergent due to the decrease,
\be\label{lnT1}
    \ln T^{\rm TE}_1(k)\sim \frac{\om_p^4}{16k^4}\,e^{2ikL}\quad \mbox{for}\quad
    k\to\infty,
\ee
which can be checked using \Ref{T2}.

Taken in the form of Eq. \Ref{Econt}, we can calculate ${E}^{\rm TE}_{\rm cont}$ numerically since it is represented by a convergent integral. For technical purposes, it is useful to turn the integration path slightly up into the   complex plane, $k\to ke^{i\al}$ with $\al\gtrsim0$. Under such substitution $ {E}^{\rm TE}_{\rm cont}$ does not change, but the integral converges much more rapidly.

\begin{SCfigure}[1.1][t]
{\vspace{0.8cm}\includegraphics[width=7.cm]{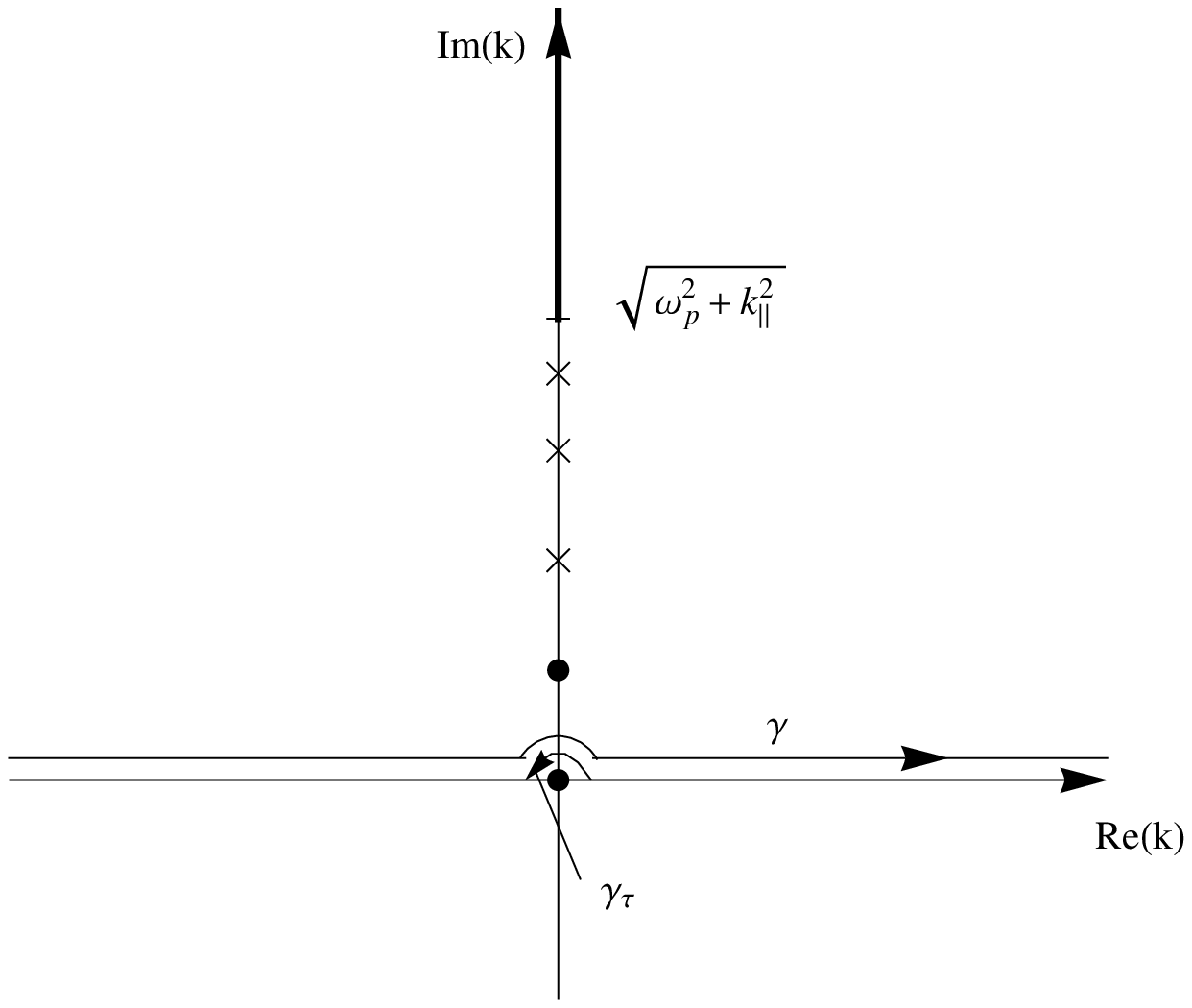}}
\caption{The complex $k$-plane.
The dots represent the poles of the integrands in \Ref{Etcont5} and \Ref{F2} resulting from zeros of the reflection coefficients, the crosses represent poles resulting from the poles of the reflection coefficients. Their locations correspond to the waveguide and surface modes whose frequencies are shown in Fig. \ref{fig2}.
The pathes $\gamma_\tau$  and $\gamma$ are used in Eqs. \Ref{Etcont5} and \Ref{F2}. On the imaginary axis there is a cut, starting from $\om_p$ in the TE case (see Eq. \Ref{Etimag1} or from $\sqrt{\om_p^2+k_{||}^2}$ (TM case (see Eq. \Ref{F5}).}\label{fig3}
\end{SCfigure}

Next we establish the relation to the representation on the imaginary frequency axis in order to get the TE part of the Lifshitz formula, \Ref{LF}. We start from  \Ref{Etcont2}, which we rewrite in the form
\be\label{Etcont3a}  \tilde{E}^{\rm TE}_{\rm cont}=
                    \frac{-1}{4\pi(3-2s)}
             \int_0^\infty\frac{dk}{2i\pi}\left(\om_p^2+k^2\right)^{\frac32-s}
             \frac{d}{dk}\ln\frac{T^{\rm TE}(k)}{T^{\rm TE}(-k)}\,,
\ee
using \Ref{Tk} and keep $s>\frac32$. We are going to change the integration path in \Ref{Etcont2} following the procedure in \cite{bord95-28-755}. The complex $k$-plane is shown in Fig. \ref{fig3}.
The integration   in \Ref{Etcont3a} is over the real $k$-axis, $k\in[0,\infty)$. On the imaginary axis, the integrand has simple poles with residua equal to (-1) at the locations $k=i\ka_j^{\rm TE}$, Eq. \Ref{kaj1}. Above, starting from $k=i\om_p$, a cut starts which results from the factor $\left(\om_p^2+k^2\right)^{3/2-s}$ in the integrand.  Furthermore we mention that $T^{\rm TE}(k)$ has a zero in $k=0$. In the quotient in  the logarithm in \Ref{Etcont3a} these zeros cancel and the integrand is regular in $k=0$.

Now we represent the logarithm in \Ref{Etcont3} as a difference, $\ln T^{\rm TE}(k)-\ln T^{\rm TE}(-k)$, and split the integral into two accordingly. In the second integral we change the variable, $k\to -k$. In doing this we pay attention to the pole in $k=0$. Therefore we write
\be\label{Etcont4} \tilde{E}^{\rm TE}_{\rm cont}=
                    \frac{-1}{4\pi (3-2s)}\lim_{\tau\to0}
                    \left(\int_{-\infty}^{-\tau}+\int_{\tau}^{\infty}\right)
                    \frac{dk}{2i\pi}\left(\om_p^2+k^2\right)^{\frac32-s}
             \frac{d}{dk} \ln{T^{\rm TE}(k)} \,,
\ee
It is our intention to move the integration path upward in the complex plane. For this, we need to close the path by adding and subtracting a path half encircling the pole in $k=0$,
\be\label{Etcont5} \tilde{E}^{\rm TE}_{\rm cont}=
                    \frac{-1}{4\pi(3-2s)}\lim_{\tau\to0}
                    \left(\int_{\gamma}+\int_{\gamma_\tau}\right)
                    \frac{dk}{2i\pi}\left(\om_p^2+k^2\right)^{\frac32-s}
             \frac{d}{dk}\ln{T^{\rm TE}(k)} \,.
\ee

The paths $\gamma_\tau$ and $\gamma$ are shown in Fig. \ref{fig3}. In the limit $\tau\to0$, the contribution from the path $\gamma_\tau$ picks up half a pole in $k=0$ and the path $\gamma$ becomes independent of $\tau$. We get
\be\label{Etcont6} \tilde{E}^{\rm TE}_{\rm cont}=
                    \frac{-1}{4\pi(3-2s)} \left[
                    \frac12 \,\om_p^{3-2s}+
                   \int_\gamma
                    \frac{dk}{2i\pi}\left(\om_p^2+k^2\right)^{\frac32-s}
             \frac{d}{dk}\ln{T^{\rm TE}(k)} \right]\,.
\ee
Now, we move the path $\gamma$ upward   to $k=i\om_p\,$. We cross the poles in $k=i\ka_j^{\rm TE}$ which delivers just the sum $\tilde{E}^{\rm TE}_{\rm wg}$, Eq. \Ref{Etwg2}, with opposite sign. After that we tighten the path around the cut on the imaginary axis, $k\to i\ka$, and with
\be\label{cut}  \left(\om_p^2+(i\ka)^2\right)^{\frac32-s}-\left(\om_p^2+(-i\ka)^2\right)^{\frac32-s}
    =-2i\cos(\pi s)\left(\ka^2-\om_p^2\right)^{\frac32-s}   \ee
we get
\be\label{Etcont7} \tilde{E}^{\rm TE}_{\rm cont}=
                    \frac{-1}{4\pi(3-2s)} \left[
                    \frac12 \,\om_p^{3-2s}-\tilde{E}^{\rm TE}_{\rm wg}\right]
                +\tilde{E}^{\rm TE}_{\rm imag},
\ee
where
\be\label{Etimag1} \tilde{E}^{\rm TE}_{\rm imag}=
                     \frac{\cos(\pi s)}{4\pi^2(3-2s)}\int_{\om_p}^\infty d\ka\, \left(\ka^2-\om_p^2\right)^{\frac32-s}
                    \frac{d}{d\ka}\,\ln T^{\rm TE}(i\ka)
\ee
is the unrenormalized vacuum energy represented as integral over the imaginary  axis. It is, however, still different from the corresponding expression in \Ref{LF} in having the transmission coefficient $ T^{\rm TE}$ instead of $ T_1^{\rm TE}$ in \Ref{LF}. To proceed, we make use of the factorization \Ref{T2} which, analytically continued to the imaginary axis, takes the form
\be\label{T4}    T^{\rm TE}(i\ka)=T_1^{\rm TE}(i\ka)\,
                \frac{4\eta\ka}{(\ka+\eta)^2}\,e^{- (\eta-\ka)L}.
\ee
For the analytic continuation of the momenta we use the notations $k=i \ka$, $q=i\eta$.
The relation $q^2=\om_p^2+k^2$, following from \Ref{D2}, translates into $\eta^2=\ka^2-\om_p^2$.
According to \Ref{T4} we split
\be\label{Etimag2}  \tilde{E}^{\rm TE}_{\rm imag}={E}^{\rm TE}_{\rm imag}+H_{\rm TE}+E_{L_2},
\ee
where, from the first factor in \Ref{T4},
\be\label{Eimag1} {E}^{\rm TE}_{\rm imag}=
                     \frac{\cos(\pi s)}{4\pi^2(3-2s)}\int_{\om_p}^\infty d\ka\, \left(\ka^2-\om_p^2\right)^{\frac32-s}
                    \frac{d}{d\ka}\,\ln T_1^{\rm TE}(i\ka),
\ee
from the second factor
\be\label{H1}   H_{\rm TE}=
                    \frac{\cos(\pi s)}{4\pi^2(3-2s)}\int_{\om_p}^\infty d\ka\, \left(\ka^2-\om_p^2\right)^{\frac32-s}
                    \frac{d}{d\ka}\,\ln  \frac{4\eta\ka}{(\ka+\eta)^2},
\ee
and from the third factor
\be\label{EL2_1}    E_{L_2}=  -\frac{\cos(\pi s)L}{4\pi^2(3-2s)}\int_{\om_p}^\infty d\ka\,
                    \left(\ka^2-\om_p^2\right)^{\frac32-s}
                    \frac{d}{d\ka}\,(\eta-\ka)
\ee
follow. The latter contribution is proportional to the separation $L$ and can be calculated easily,
\be\label{EL2_2} E_{L_2}=-\frac{\om_p^{2(2-s)}L}{12\pi^2}\,h_2(s).
\ee
The function $h_2(s)$ can be expressed in terms  of hypergeometric functions. It has a pole in $s=0$,
\be\label{h2p}     h_2(s)=\frac{-3}{16s}+O(1),
\ee
with the same residuum as $h_1(s)$, Eq. \Ref{h1p}. In the following, we will need only the relation
\be\label{h12}  \lim_{s\to0}\left(h_1(s)-h_2(s)\right)=-\frac14  \,.
\ee
Now we consider $E^{\rm TE}_{\rm imag}$, Eq. \Ref{Eimag1}. First, we note that the function
\be\label{TE1imag} T_1^{\rm TE}(i\ka)=\frac{1}{1-r_{\rm TE}^2e^{-2\eta L}},
\ee
with $r_{\rm TE}$ given by Eq. \Ref{r}, is just the same as in the first logarithm in \Ref{LF}. Since it is exponentially decreasing for $\ka\to\infty$, we can remove the regularization by simply putting $s=0$. Finally, we perform a substitution of variables, $\xi=\sqrt{\ka^2-\om_p^2}\,$,
\be\label{Eimag2}    {E}^{\rm TE}_{\rm imag}=
                    \frac{1}{12\pi^2}\int_0^\infty d\xi\,\xi^3
                    \frac{d}{d\xi}\,\ln T_1^{\rm TE}(i\ka),
\ee
where
\be\label{T5}       T_1^{\rm TE}(i\ka)=
                \frac{1}{1-\left(\frac{\ka-\xi}{\ka+\xi}\right)^2e^{-2\xi L}} \,
\ee
is the transmission coefficient on the imaginary axis appearing in \Ref{LF} after carrying out the integration over $k_{||}$ .
Expression \Ref{Eimag2} can be considered final. Because of the convergence, it is also possible to integrate by parts,
\be\label{Eimag3}    {E}^{\rm TE}_{\rm imag}=-
                    \frac{1}{4\pi^2}\int_0^\infty d\xi\,\xi^2
                    \,\ln T_1^{\rm TE}(i\ka).
\ee
Eqs. \Ref{Eimag2} and \Ref{Eimag3} are variants of representing the well--known  TE part of the vacuum energy calculated on the imaginary frequency axis. We mention that these differ from $\tilde{E}^{\rm TE}_{\rm imag}$, which we initially got from moving the integration path, by the addendum $H_{\rm TE}$, which does not depend on the separation $L$, and by $E_{L_2}$, which is proportional to $L$. In the sense of a renormalization it is natural to remove these two contributions and to consider ${E}^{\rm TE}_{\rm imag}$ as the renormalized vacuum energy. Of course, ${E}^{\rm TE}_{\rm imag}$ has the necessary behavior for large separation,
\be\label{Eimag4}  {E}^{\rm TE}_{\rm imag}\to
                -\frac{\pi^2}{1440 L^3}\quad\mbox{for}\quad L\to\infty\,.
\ee
It is also easy to obtain the behavior for small separation. One can put  $L=0$ in the integrand in \Ref{Eimag2} since the remaining integral converges. A short calculation gives
\be\label{Eimag5}  {E}^{\rm TE}_{\rm imag}\to
                -\frac{3\pi-8}{72\pi^2}\,\om_p^3\quad\mbox{for}\quad L\to 0\,.
\ee
This limit coincides with that found in \cite{bord95-28-755}, Eq. (41). In the complete energy, this contribution is subleading. The leading contributions comes from the TM polarization, Eq. \Ref{MEimag6}.

It remains to consider $H_{\rm TE}$, Eq. \Ref{H1}. The integration can be carried out explicitly,
\be\label{H2}   H_{\rm TE}=\frac{\cos(\pi s)}{4\pi(3-2s)}\left(\frac12-\frac{4}{3\pi}\right)\,\om_p^{3-2s}.
\ee
With this, we return to $\tilde{E}^{\rm TE}_{\rm cont}$, Eq. \Ref{Etcont7}, and,  inserting from \Ref{Etimag2} and \Ref{H2}, we get
\be\label{Etcont8} \tilde{E}^{\rm TE}_{\rm cont}=
                    -\frac{\cos(\pi s)}{3(3-2s)\pi^2}\,\om_p^{3-2s}+
                    \tilde{E}^{\rm TE}_{\rm imag}+E_{L_2}\,.
\ee
The contribution from the half pole in $k=0$ canceled. We mention that we have to keep $s>\frac32$ (in fact, $s>0$ is sufficient) in this expression because of the pole in $E_{L_2}$.

The equations \Ref{Etcont8} and \Ref{Et1}, together with \Ref{Etwg1} and \Ref{Etcont3}, are two representations for $\tilde{E}^{\rm TE}$, one with integration on the imaginary axis, and the other with summation and integration on the real axis,
\bea\label{Et2} \tilde{E}^{\rm TE}&=&
            -\frac{\cos(\pi s)}{3(3-2s)\pi^2}\,\om_p^{3-2s}+
                    {E}^{\rm TE}_{\rm imag}+E_{L_2}
\nn\\            &=&\tilde{E}^{\rm TE}_{\rm wg}+ {E}^{\rm TE}_{\rm cont}+E_{L_1}\,.
\eea
Needless to stress that these are equal, at least for $s>\frac32$. Now we perform the renormalization by subtracting from both sides the first two contributions in the first line. The first of these   does not depend on $L$ and the other, $E_{L_2}$,   is proportional to $L$. Thus we define
\be\label{ETE1} E^{\rm TE}=\tilde{E}^{\rm TE}+\frac{\cos(\pi s)}{3(3-2s)\pi^2}\,\om_p^{3-2s}-E_{L_2}
\ee
as the renormalized vacuum energy. It has all necessary properties since from \Ref{Et2} and \Ref{ETE1}
\be\label{ETE2} E^{\rm TE}={E}^{\rm TE}_{\rm imag}
\ee
follows by definition, for which these properties are known.

Now we insert the second line from \Ref{Et2} into \Ref{ETE1} and come to
\be\label{ETE3}     E^{\rm TE}=E^{\rm TE}_{\rm wg}+E^{\rm TE}_{\rm cont},
\ee
where $E^{\rm TE}_{\rm cont}$ is given by Eq. \Ref{Econt}, and where we defined
\be\label{ETEwg} E^{\rm TE}_{\rm wg}=\tilde{E}^{\rm TE}_{\rm wg}+
        \frac{\cos(\pi s)}{3(3-2s)\pi^2}\,\om_p^{3-2s}+E_{L_1}-E_{L_2}.
\ee
Now we are finally in a position to remove the regularization. The pole in $s=0$ is present only in the last two terms in \Ref{ETEwg} and cancels. Using \Ref{h12} together with \Ref{EL1_2} and \Ref{EL2_2}, we  get for $s=0$
\be\label{ETEwg2} E^{\rm TE}_{\rm wg}=
            -\frac{1}{12\pi}\sum_j\left(q_j^{\rm TE}\right)^3+\frac{\om_p^3}{9\pi^2}+\frac{\om_p^4L}{48\pi^2}\,.
\ee
We included the contribution proportional to $L$, resulting from $E_{L_1}$ and $E_{L_2}$, and the constant contribution, resulting from $H$, into the waveguide contribution for the following reason. It can be shown, that the sum in \Ref{ETEwg2} has for large $L$ an asymptotic behavior
\be\label{ETwg4}    \tilde{E}^{\rm TE}_{\rm wg}=
                    -\frac{\om_p^4L}{48\pi^2}  -  \frac{\om_p^3}{9\pi^2}+\dots\,,
\ee
where the dots denote bounded, but oscillating contributions.

In \Ref{ETEwg2}, the first two terms are just subtracted. Hence, $E^{\rm TE}_{\rm wg}$, Eq. \Ref{ETEwg2}, at most oscillates for $L\to\infty$, which justifies    considering it as the contribution from the waveguide modes to the vacuum energy after renormalization.
\begin{figure}[t]
\includegraphics[width=14.cm]{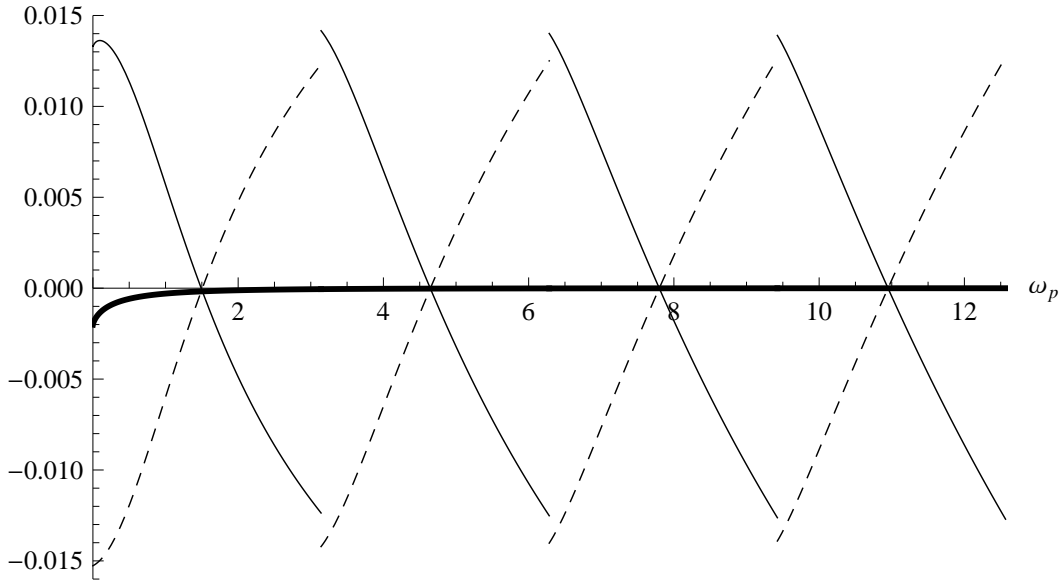}
\caption{The contributions to the vacuum energy for the TE polarization as function of $\om_p$ for $L=1$.
The solid line represents the vacuum energy, ${E}^{\rm TE}_{\rm imag}={E}^{\rm TE}_{\rm wg}+{E}^{\rm TE}_{\rm cont}$, the thin solid line represents the contribution from the photonic modes, ${E}^{\rm TE}_{\rm cont}$,  and the dashed line is the contribution from the waveguide modes, ${E}^{\rm TE}_{\rm wg}$. The curves are equivalent to represent the energies multiplied by $L^3$ as function of $L$ for $\om_p=1$. The jumps appear when $\om_p$ goes through multiples of $\pi$.}\label{figTE}
\end{figure}

Collecting from \Ref{ETE2} and \Ref{ETE3}, we have with
\be\label{ETE4}     {E}^{\rm TE}={E}^{\rm TE}_{\rm imag}={E}^{\rm TE}_{\rm wg}+{E}^{\rm TE}_{\rm cont}
\ee
two representations of the renormalized vacuum energy resulting from the TE polarization, one with integration over the imaginary axis, which coincides with the corresponding part in the Lifshitz formula, and another one with summation and integration over real frequencies, i.e., over the spectrum. Both sides can be evaluated numerically. Both must deliver the same numbers. Indeed, they do. The results are represented in Fig. \ref{figTE} as a function of $\om_p$ with $L=1$.
It must be mentioned that, already for dimensional reasons, the energy can be represented in the form
\be\label{ETE5} {E}^{\rm TE}=\frac{1}{L^3}\,f(\om_p\,L),
\ee
where $f(\om_p\,L)$ is dimensionless (its limiting values can be read off from  Eqs. \Ref{Eimag4} and \Ref{Eimag5}). Thus a plot of ${E}^{\rm TE}$ as function of $\om_p$ is equivalent to a plot of ${E}^{\rm TE}L^3$ as function of $L$.

In Fig. \ref{figTE} also ${E}^{\rm TE}_{\rm wg}$ and ${E}^{\rm TE}_{\rm cont}$ are shown as functions of $\om_p$. These show an oscillating behavior with strong compensation in a way  that their sum is just the much smaller  ${E}^{\rm TE}_{\rm imag}$, shown as thick solid line.

\subsection{TM case}
In this subsection we consider the TM part of the vacuum energy. We start from the unrenormalized one, $\tilde{E}^{\rm TM}$, in Eq. \Ref{EtTETM}. According to the spectrum of modes described in Section 2, it is a sum,
\be\label{MEt1} \tilde{E}^{\rm TM}=
                \tilde{E}_{\rm sf}+\tilde{E}^{\rm TM}_{\rm wg}+\tilde{E}^{\rm TM}_{\rm cont},
\ee
of surface modes,
\be\label{MEtsf1}   \tilde{E}_{\rm sf}=
                \frac12\int\frac{d\mathbf{k}_{||}}{(2\pi)^2}\, e^{-\delta k_{||}}
                \sum_{sf}\om_{sf}^{1-2s},
\ee
of waveguide modes,
\be\label{MEtwg1}   \tilde{E}_{\rm wg}^{\rm TM}=
                \frac12\int\frac{d\mathbf{k}_{||}}{(2\pi)^2}\, e^{-\delta k_{||}}
                \sum_{j}\left(\om_{j}^{\rm TM}\right)^{1-2s},
\ee
and of photonic modes,
\be\label{MEtcont1}   \tilde{E}_{\rm cont}^{\rm TM}=
                \frac12\int\frac{d\mathbf{k}_{||}}{(2\pi)^2}\, e^{-\delta k_{||}}
                \int_0^\infty\frac{dk}{2i\pi}\left(\om_p^2+k_{||}^2+k^2\right)^{\frac12-s}
                \frac{d}{dk}\ln\frac{T^{\rm TM}(k)}{\left(T^{\rm TM}(k)\right)^*}\,.
\ee
It must be mentioned that we are forced to introduce an additional regularization for the integration over the momentum $\mathbf{k}_{||}$. The reason  is in the boundedness of the frequencies of the surface modes, as expressed by Eq. \Ref{sfb}, which makes the zetafunctional regularization effectless. This is some kind of additional divergence which was not observed earlier. However, it does not show up in the force. As we will see below, all terms divergent for $\delta\to0$, do not depend on the width $L$ of the gap.

The main problem, we are faced with, is to perform the analytic continuation in the zetafunctional parameter $s$ to $s=0$. In the TE case this was simple since, after performing the integration over $\mathbf{k}_{||}$, we could perform the analytic continuation explicitly.  In the TM case we cannot integrate over $\mathbf{k}_{||}$ in such simple way since it enters the reflection coefficient. However, below we will show how this problem can be solved.

We start with the surface mode contribution \Ref{MEtsf1}. Here we can make use of the circumstance discussed at the end of Section 2, that the frequencies, for large $k_{||}$, approach the frequency $\om_{single}$, \Ref{omsi}, of a surface plasmon on a single  interface. Therefore we subtract  $\om_{single}$ and add it back,
\be\label{MEtsf2}   \tilde{E}_{\rm sf}= {E}_{\rm sf} +2E_{single}\,,
\ee
where
\be\label{MEsf1} {E}_{\rm sf}=\frac12\int\frac{d\mathbf{k}_{||}}{(2\pi)^2}\, e^{-\delta k_{||}}
                \sum_{sf}
                \left[\left(\om_{sf}\right)^{1-2s}-\left(\om_{single}\right)^{1-2s}\right] ,
\ee
is the subtracted energy of the surface modes  and
\be\label{MEsi1} E_{single}=\frac12\int\frac{d\mathbf{k}_{||}}{(2\pi)^2}\, e^{-\delta k_{||}}
                \sum_{sf}
               \left(\om_{single}\right)^{1-2s}
\ee
is the energy of a plasmon on a single interface. In the energy ${E}_{\rm sf}$ one can remove the regularizations, i.e., one can put $s=0$ and $\delta=0$, and the emerging integral turns  out to be    convergent. The integral in $E_{single}$, Eq. \Ref{MEsi1} is quite simple and can be calculated explicitly for $s=0$ and $\delta\to0$,
\be\label{MEsi2} E_{single}=
    \left(\frac{1-12\ln2+6\ln\om_p}{192\sqrt{2}\pi}
    +\frac{\ln(\delta \om_p)+\gamma)}{32\sqrt{2}\pi}\right)\om_p^3
    +\frac{1}{4\sqrt{2}\pi}\frac{\om_p}{\delta^2}+O(\delta),
\ee
where $\gamma$ is the Euler constant.
It has two terms divergent for $\delta\to0$, one quadratic and one logarithmic. We mention that $ E_{single}$ does not depend (by construction) on the separation $L$ and therefore all these terms depend on $\om_p$ only.

Next we turn to the waveguide mode contribution, Eq. \Ref{MEtwg1}. First, we mention that we can  remove there the additional regularization putting $\delta=0$. This is because the frequencies $\om_j^{\rm TM}$ of the  waveguide modes grow proportional to $k_{||}$ for $k_{||}\to\infty$ and the zetafunctional regularization is effective here.
In order to perform the analytic continuation in $s$,  we subtract  the frequencies of the TE waveguide modes and add them back,
\be\label{MEtwg2}  \tilde{E}_{\rm wg}^{\rm TM}=
                \Delta{E}_{\rm wg}+\tilde{E}_{\rm wg}^{\rm TE},
\ee
where
\be\label{dEwg1}  \Delta{E}_{\rm wg}=
            \frac12\int\frac{d\mathbf{k}_{||}}{(2\pi)^2}\,
                \sum_{j}\left[\left(\om_{j}^{\rm TM}\right)^{1-2s}-
                    \left(\om_{j}^{\rm TE}\right)^{1-2s}\right]
\ee
has the differences of the frequencies and $\tilde{E}_{\rm wg}^{\rm TE}$ is the TE contribution \Ref{Etwg1}. The latter has been treated  in the preceding subsection and the analytic continuation in $s$ is given by Eq, \Ref{Etwg2}. The difference energy, $\Delta{E}_{\rm wg}$, can be continued in $s$ simply by putting $s=0$ in \Ref{dEwg1} since both frequencies approach each other for $k_{||}\to \infty$, as mentioned after Eq. \Ref{omj4}.

Finally, we consider the energy $ \tilde{E}_{\rm cont}^{\rm TM}$, \Ref{MEtcont1}, of the photonic modes in \Ref{MEt1}. We proceed similar as just above and subtract  the corresponding TE frequencies,
\be\label{MEtcont2}     \tilde{E}^{\rm TM}_{\rm cont}=
                        \Delta E_{\rm cont}+ \tilde{E}^{\rm TE}_{\rm cont},
\ee
where $\tilde{E}^{\rm TE}_{\rm cont}$ is given by Eq. \Ref{Etcont1} and its analytic continuation by the subsequent equations in the preceding subsection. The difference energy,
\bea\label{dEcont1}   \Delta {E}_{\rm cont} &=&
                \frac12\int\frac{d\mathbf{k}_{||}}{(2\pi)^2}\, e^{-\delta k_{||}}
                \int_0^\infty\frac{dk}{2i\pi}\left(\om_p^2+k_{||}^2+k^2\right)^{\frac12-s}
 \nn\\&&\qquad               \frac{d}{dk}        \left[\ln\frac{T^{\rm TM}(k)}{\left(T^{\rm TM}(k)\right)^*}
                            -\ln\frac{T^{\rm TE}(k)}{\left(T^{\rm TE}(k)\right)^*}\right],
\eea
has the property that on can put $s=0$ directly since the remaining integrals do converge. Also we can put $\delta=0$ since dangerous terms cancel in the ratios inside the logarithms. For numerical evaluation we used
\be\label{dEcont2}   \Delta {E}_{\rm cont} =
                \frac{1}{4\pi^2}\int_0^\infty dk\,
                \int_0^\infty dk_{||} k_{||}\left(\om_p^2+k_{||}^2+k^2\right)^{\frac12}
               \frac{d}{dk}    \,    {\rm Im} \left(
                            \ln\frac{T^{\rm TM}(k)}{T^{\rm TE}(k)}  \right),
\ee
which differs from \Ref{dEcont1} by simple rewriting, including a change in the order of the integrations.

Now, after having clarified how to make the analytic continuation in the contributions from the real frequencies, we turn to the imaginary frequencies. Again, we follow the procedure used in the preceding subsection. We are going to transform the integration path of the $k$ integration. We do this for fixed $k_{||}$. For that reason, it is useful to start from the contribution of the photonic modes, $\tilde{E}_{\rm cont}^{\rm TM}$, \Ref{MEtcont1},  which we represent in the form 
\be\label{MEtcont3} \tilde{E}_{\rm cont}^{\rm TM}=
                         \frac12\int\frac{d\mathbf{k}_{||}}{(2\pi)^2} \, e^{-\delta k_{||}}
                         F(k_{||}),
\ee
with, using \Ref{Tk},
\be\label{F1} F(k_{||})=\int_0^\infty\frac{dk}{2i\pi}\left(\om_p^2+k_{||}^2+k^2\right)^{\frac12-s}
                \frac{d}{dk}\ln\frac{T^{\rm TM}(k)}{T^{\rm TM}(-k)}.
\ee
The transmission coefficient $T^{\rm TM}(k)$ has properties similar to $T^{\rm TE}(k)$, except for an additional zero for $k=ik_{||}$.  For instance, it has a zero in $k=0$. So, from the ratio inside the logarithm,  we split the logarithm as a difference of two and, further, the integral as a difference of two with substitution $k\to -k$ in the second. Accounting for the pole of the integrand in $k=0$ we come to
\be\label{F2} F(k_{||})=
                   \lim_{\tau\to0}
                    \left(\int_{-\infty}^{-\tau}+\int_{\tau}^{\infty}\right)
                    \frac{dk}{2i\pi}\left(\om_p^2+k^2\right)^{\frac32-s}
             \frac{d}{dk} \ln{T^{\rm TM}(k)} \,.
\ee
We proceed by closing the integration path around $k=0$ as explained in the TE case resulting in
\be\label{F3} F(k_{||})=
                   \lim_{\tau\to0}
                     \left(\int_{\gamma}+\int_{\gamma_\tau}\right)
                    \frac{dk}{2i\pi}\left(\om_p^2+k^2\right)^{\frac32-s}
             \frac{d}{dk} \ln{T^{\rm TM}(k)} \,.
\ee
Now, as mentioned, we have an additional pole in $k=ik_{||}$. It may be located inside the contour $\gamma_\tau$ or outside in dependence on $k_{||}$. It can be seen, that the case, when this pole   is located inside, results in a contribution to \Ref{MEtcont3} having an integration over $\mathbf{k}_{||}$ with $k_{||}\le\tau$  which has for $\tau\to0$ an additional smallness $\sim \tau^2$ and the corresponding contribution vanishes in the limit. So we are left with the case that this pole is outside. Then the contribution from $\gamma_\tau$ can be calculated directly and, as before, $\gamma$ becomes independent on $\tau$, and we get

\be\label{F4} F(k_{||})=\frac12 \left(\om_p^2+k_{||}^2\right)^{\frac12-s}+
          \int_{\gamma}
                    \frac{dk}{2i\pi}\left(\om_p^2+k_{||}^2+k^2\right)^{\frac32-s}
             \frac{d}{dk} \ln{T^{\rm TM}(k)} \,.
\ee
Now, we move the path upward until $k=i\sqrt{\om_p^2+k_{||}^2}$, crossing  poles of the integrand. These result from the poles of $T^{\rm TM}(k)$ in the points $k=i\ka_{sf}$ (corresponding to the surface modes) and from the poles of $T^{\rm TM}(k)$ in the points $k=i\ka_{j}$ (corresponding to the waveguide modes) having residua with negative sign. Further we have a pole from the zero of  $T^{\rm TM}(k)$ in $k=ik_{||}$ (which was not present in the TE case). Finally, we tighten the path around the cut starting now from $k=i\sqrt{\om_p^2+k_{||}^2}$ and come to the representation
\bea\label{F5}    F(k_{||})&=& \frac12 \left(\om_p^2+k_{||}^2\right)^{\frac12-s}+
                \om_p^{1-2s}-\sum_{sf}\om_{sf}^{\frac12-s}
                            -\sum_j \left(\om_j^{\rm TM}\right)^{\frac12-s}
 \nn\\&&                           +\frac{\cos(\pi s)}{\pi}\int_{\sqrt{\om_p^2+k_{||}^2}}^\infty d\ka
                            \left(\ka^2-\om_p^2-k_{||}^2\right)^{\frac12-s}
                            \frac{d}{d\ka}\,\ln T^{\rm TM}(i\ka).
\eea
We mention that the cosine appears with opposite sign as compared to \Ref{cut} because of the different number in front of $s$ in the exponential.

Now we insert \Ref{F5} into \Ref{MEtcont3}. The integration over $k_{||}$ can be carried out in the first two contributions, for the  next two we use \Ref{MEtsf1} and \Ref{MEtwg1} and come to
\be\label{MEtcont4} \tilde{E}^{\rm TM}_{\rm cont}
                    =-\frac{\om_p^3}{24\pi}+\frac{1}{4\pi^2}\frac{\om_p}{\delta^2}
                    -\tilde{E}_{sf}-\tilde{E}^{\rm TM}_{\rm wg}+\tilde{E}^{\rm TM}_{\rm imag},
\ee
where we defined
\be\label{MEtimag1} \tilde{E}^{\rm TM}_{\rm imag}=
                        \frac{\cos(\pi s)}{2\pi}
                        \int\frac{d\mathbf{k}_{||}}{(2\pi)^2}\, e^{-\delta k_{||}}
                         \int_{\sqrt{\om_p^2+k_{||}^2}}^\infty d\ka
                         \left(\ka^2-\om_p^2-k_{||}^2\right)^{\frac12-s}
                            \frac{d}{d\ka}\,\ln T^{\rm TM}(i\ka).
\ee
The latter is the unrenormalized vacuum energy represented as an integral over the imaginary axis. It differs from the corresponding contribution in \Ref{LF} by the different transmission coefficients, $T^{\rm TM}(i\ka)$ here and $T^{\rm TM}_1(i\ka)$, Eq. \Ref{T3}, in \Ref{LF}.

Now we consider this difference using \Ref{T2}, analytically continued to the imaginary axis, where it takes the form
\be\label{T5a}    T^{\rm TM}(i\ka)=T_1^{\rm TM}(i\ka)\,
                \frac{4\ep(i\xi)\eta\ka}{(\ep(i\xi)\ka+\eta)^2}\,e^{- (\eta-\ka)L}.
\ee
We use the same notations as in Eq. \Ref{T4} and, in addition, the analytic continuation, $\om=i\xi$, of the frequency, resulting in the relations
 $\xi=\sqrt{\ka^2-\om_p^2-k_{||}^2}$ and $\ep(i\xi)=1+\frac{\om_p^2}{\xi^2}=\frac{\ka^2-k_{||}^2}{\ka^2-k_{||}^2-\om_p^2}$ ,
 which must be used in \Ref{T5a}.

According to the factors in \Ref{T5a}, we split,
\be\label{MEtimag2} \tilde{E}^{\rm TM}_{\rm imag}={E}^{\rm TM}_{\rm imag}+H_{\rm TM}+E_{L_2},
\ee
where $E_{L_2}$ is the same as in the TE case, Eq. \Ref{EL2_1} with its analytic continuation \Ref{EL2_2}.
In the first contribution, we can remove the regularization since the integrations are convergent such that it takes the form
\be\label{MEimag2} {E}^{\rm TM}_{\rm imag}=
     \frac{1}{2\pi}
                        \int\frac{d\mathbf{k}_{||}}{(2\pi)^2}\,
                         \int_{\sqrt{\om_p^2+k_{||}^2}}^\infty d\ka
                         \left(\ka^2-\om_p^2-k_{||}^2\right)^{\frac12}
                            \frac{d}{d\ka}\,\ln T_1^{\rm TM}(i\ka).
\ee
The reflection coefficient on the imaginary axis is
\be\label{TM1imag} T_1^{\rm TM}(i\ka)=\frac{1}{1-r_{\rm TM}^2e^{-2\eta L}},
\ee
with $r_{\rm TM}$ given by Eq, \Ref{r}. Thus, Eq. \Ref{MEimag2} is just the TM contribution in the Lifshitz formula \Ref{LF}.
It can be rewritten also by making a change of the integration variable $\ka$ for $\xi$,
\be\label{MEimag3} {E}^{\rm TM}_{\rm imag}=
     \frac{ 1}{4\pi^2}
                        \int_{0}^\infty d{k}_{||}
                         \int_{0}^\infty d\xi   \, \xi
                            \frac{d}{d\xi}\,\ln T_1^{\rm TM}(i\ka),
\ee
or, by integrating by parts,  in the form%
\be\label{MEimag4} {E}^{\rm TM}_{\rm imag}=
     \frac{-1}{4\pi^2}
                        \int_{0}^\infty d{k}_{||}\,k_{||}
                         \int_{0}^\infty d\xi   \,  \ln T_1^{\rm TM}(i\ka).
\ee
In \Ref{MEimag3} and \Ref{MEimag4} the  variables must be expressed according to
\bea\label{intermsofxi} \ka&=&\sqrt{\xi^2+\om_p^2+k_{||}^2},   \nn \\
                        \eta&=&\sqrt{k_{||}^2+\xi^2},        \nn\\
                        \ep(i\xi)&=&1+\frac{\om_p^2}{\xi^2}.
\eea
For later use we define the difference between the corresponding TM and TE contributions,
\be\label{dEimag1}  \Delta E_{\rm imag}={E}^{\rm TM}_{\rm imag}-{E}^{\rm TE}_{\rm imag}.
\ee
The TM contributions has the following asymptotic properties. For large separation $L$, or, equivalently, for large $\om_p$, it turns into the ideal conductor expression,
\be\label{MEimag5}  {E}^{\rm TM}_{\rm imag}\to
                -\frac{\pi^2}{1440 L^3}\quad\mbox{for}\quad L\to\infty\,,
\ee
in parallel to \Ref{Eimag4}. For small separation, or small $\om_p$, we make in \Ref{MEimag4} the substitution $\xi\to \om_p\xi$. After that, one can put $\om_p=0$ in the integrand and come to
\be\label{MEimag6}  {E}^{\rm TM}_{\rm imag}\to
                -c_2\frac{\om_p}{L^2},
\ee
where the remaining integral,
\be\label{c2}   c_2=\frac{1}{4\pi^2} \int_{0}^\infty d{k}_{||}\,k_{||}
                         \int_{0}^\infty d\xi   \,
                         \ln \left(1-\frac{e^{-2k_{||}}}{(1+2\xi^2)^2}\right)^{-1},
\ee
is a number, $c_2 \simeq0.00391$, which coincides with that found in \cite{bord95-28-755}, Eq. (41) and earlier in Eq. (14) in \cite{lamb00-8-309} and in \cite{gene04-29-311}, where a sum representation was derived. This expression can also be calculated starting from the contribution of the surface modes, Eq. \Ref{MEsf2}. The corresponding calculation is shown in Appendix B.
We mention that Eq. \Ref{MEimag6} with \Ref{c2}   is the same as \Ref{vdW} with \Ref{ep1} inserted.

Finally, the contribution
\be\label{HTM} H_{\rm TM}=\frac{\cos(\pi s)}{2\pi}
                        \int\frac{d\mathbf{k}_{||}}{(2\pi)^2}\, e^{-\delta k_{||}}
                      \int_{0}^\infty d\xi\,
                         \xi^{1-2s}\frac{d}{d\ka}\,
                          g(k_{||},\xi)
\ee
in \Ref{MEtimag2}, resulting from the middle factor in \Ref{T5a}, needs to be considered. In \Ref{HTM} we changed the integration over $\ka$ for $\xi$ (\Ref{intermsofxi} must be used) and introduced the notation
\be\label{g1}  g(k_{||},\xi) = \ln\frac{4\ep(i\xi)\eta\ka}{(\ep(i\xi)\ka+\eta)^2}.
\ee
To proceed, we represent this function as a sum of three,
\be\label{g2} g(k_{||},\xi) = g_A +g_B +g_{\rm TE},
\ee
where
\bea\label{g} g_A&=&\ln \frac{4(\om_p^2+\xi^2)\xi^2}{(\om_p^2+2\xi^2)^2},
\nn \\       g_B&=&2\ln \frac{4(\om_p^2+2\xi^2)(\ka+\eta)}{2\xi^2(\ka+\ep(i\xi) \eta)},
\nn\\          g_{\rm TE}&=&\ln \frac{4\eta\ka}{(\eta+\ka)^2}.
\eea
According to the splitting \Ref{g2}, we represent \Ref{HTM} as a sum,
\be\label{H3}    H_{\rm TM}=H_A+H_B+H_{\rm TE}
\ee
The last term, following from the function  $g_{\rm TE}$, is the same as appeared   in the TE case, Eq.\Ref{H1} and can be treated in the same way as in the preceding subsection. The result is given by Eq. \Ref{H2}.

Now we consider $H_A$. Since $g_A$ depends on $\xi$ only, the integrations can be carried out easily. For $s=0$ and $\delta\to0$ we get
\be\label{HA}   H_A=\frac{\sqrt{2}-1}{4\pi}\frac{\om_p}{\delta^2}+O(\delta).
\ee
This term has a quadratic divergence in the regularization parameter $\delta$. This is, after $\tilde{E}_{sf}$ (see Eqs. \Ref{MEtsf2} and \Ref{MEsi2}), the second place  where it appears.

Finally, we   consider $H_B$,
\be\label{HB1}  H_B=\frac{\cos(\pi s)}{2\pi}
                        \int\frac{d\mathbf{k}_{||}}{(2\pi)^2}\, e^{-\delta k_{||}}
                      \int_{0}^\infty d\xi\,
                         \xi^{1-2s}\frac{d}{d\ka}\,
                          g_B \,,
\ee
with $g_B$ given by the second line in \Ref{g}. In order to perform the continuation to $s=0$ and to $\delta=0$, we subtract and add back the asymptotics of $g_B$ for $k_{||}\to\infty$,
\be\label{g3}g_B\sim\frac{1}{2(\om_p^2+2\xi^2)(\om_p^2+k_{||}^2)},
\ee
splitting $H_B$ into two parts,
\be\label{HB2}H_B=H_{B_1}+H_{B_2}.
\ee
In the first part, $H_{B_1}$, the integrations converge for $s=0$ and $\delta=0$,
\be\label{HB3}H_{B_1}=\frac{1}{2\pi^2}
                        \int\frac{d\mathbf{k}_{||}}{(2\pi)^2}\, 
                      \int_{0}^\infty d\xi\,
                         \xi\frac{d}{d\ka}\,
                          \left(g_B-\frac{1}{2(\om_p^2+2\xi^2)(\om_p^2+k_{||}^2)}\right) \,,
\ee
These integrations can be carried out numerically, resulting in
\be\label{HB4}H_{B_1}=\frac{c}{2\pi^2}\,\om_p^3
\ee
with $c\simeq 0.06777$. In the second part,
\be\label{HB5}H_{B_2}=\frac{1}{2\pi^2}
                        \int\frac{d\mathbf{k}_{||}}{(2\pi)^2}\, 
                      \int_{0}^\infty d\xi\,
                         \xi\frac{d}{d\ka}\,
                         \frac{1}{2(\om_p^2+2\xi^2)(\om_p^2+k_{||}^2)} \,,
\ee
the integration can be carried out analytically in the limit $\delta\to0$ for $s=0$,
\be\label{HB6}H_{B_2}=\frac{(\ln(\om_p\,\delta)+\gamma)}{16\sqrt{2}\pi}\,\om_p^3+O(\delta),
\ee
where $\gamma$ is the Euler constant.
Together, from \Ref{HB4} and \Ref{HB6}, we get
\be\label{HB7}H_B=\left(\frac{c}{2\pi^2}+\frac{(\ln(\om_p\,\delta)+\gamma)}{16\sqrt{2}\pi}\right)
        \om_p^3+O(\delta)\,.
\ee
This must be inserted, together with  \Ref{HA} and \Ref{H2}, into \Ref{H3},
\be\label{HTM2} H_{\rm TM}=
        \left(\frac{1}{24\pi}-\frac{1}{9\pi^2}+
        \frac{c}{2\pi^2}+\frac{(\ln(\om_p\,\delta)+\gamma)}{16\sqrt{2}\pi}\right)
        \om_p^3+\frac{\sqrt{2}-1}{4\pi}\frac{\om_p}{\delta^2}+O(\delta)\,,
\ee
which completes the analytic continuation in $\tilde{E}^{\rm TM}_{\rm imag}$, Eq. \Ref{MEtimag2}. Now we insert this expression into $\tilde{E}^{\rm TM}_{\rm cont}$, Eq. \Ref{MEtcont4} and obtain
\be\label{MEtcont5} \tilde{E}^{\rm TM}_{\rm cont}=- \tilde{E}_{\rm sf}-\tilde{E}^{\rm TM}_{\rm wg}+
            \left(\frac{c}{2\pi^2}-\frac{1}{9\pi^2}+
            \frac{(\ln(\om_p\,\delta)+\gamma)}{16\sqrt{2}\pi}\right)\om_p^3+
            E^{\rm TM}_{\rm imag}+E_{L_2},
\ee
whereby four terms canceled. The last step in this sequence of calculations is to insert this expression for $\tilde{E}^{\rm TM}_{\rm cont}$ into \Ref{MEt1},
\be\label{MEt2} \tilde{E}^{\rm TM}= \left(\frac{c}{2\pi^2}-\frac{1}{9\pi^2}+
            \frac{(\ln(\om_p\,\delta)+\gamma)}{16\sqrt{2}\pi}\right)\om_p^3+
            E^{\rm TM}_{\rm imag}+E_{L_2},
\ee
whereby the contributions from the surface and waveguide modes canceled.

With Eq. \Ref{MEt2}, the transition from the summations and integration over real frequencies to the integration over imaginary frequencies, including the analytic continuations, is finished. Now we can define the renormalized vacuum energy resulting from the TM polarization. We proceed similar to the TE case and define
\be\label{ETM1} E^{\rm TM}=\tilde{E}^{\rm TM}-\left(\frac{c}{2\pi^2}-\frac{1}{9\pi^2}+
            \frac{(\ln(\om_p\,\delta)+\gamma)}{16\sqrt{2}\pi}\right)\om_p^3-E_{L_2}.
\ee
This coincides, by definition, with the vacuum energy $E^{\rm TM}_{\rm imag}$ on the imaginary axis,
\be\label{ETM2} E^{\rm TM}=E^{\rm TM}_{\rm imag}.
\ee
As already mentioned,  $E^{\rm TM}_{\rm imag}$ is just the TM contribution to the Lifshitz formula \Ref{LF}.

In order to get the corresponding expression on the real axis, we insert for $\tilde{E}^{\rm TM}$ in \Ref{ETM1} from \Ref{MEt1}. We get
\be\label{ETM3}E^{\rm TM}=E^{\rm TM}_{sf}+E^{\rm TM}_{\rm wg}+E^{\rm TM}_{\rm cont},
\ee
where we defined
\be\label{ETMcont1}E^{\rm TM}_{\rm cont}=\tilde{E}^{\rm TM}_{\rm cont}-E_{L_1}.
\ee
Using \Ref{MEtcont2} and \Ref{Etcont3},  this can be rewritten as,
\be\label{ETMcont2}E^{\rm TM}_{\rm cont}=E^{\rm TE}_{\rm cont}+\Delta E_{\rm cont},
\ee
and represents the renormalized contribution from the TM photonic modes. It is represented as sum of the corresponding TE modes and the difference between TM and TE mode's  contribution.

The first contribution in $E^{\rm TM}$, Eq. \Ref{ETM3}, is defined as
\be\label{ETMsf1}   E_{sf}=\tilde{E}^{\rm TM}_{sf}-\frac{1}{2\sqrt{2}\pi}\frac{\om_p}{\delta^2}-
    \left(\frac{(\ln(\om_p\,\delta)+\gamma)}{16\sqrt{2}\pi} +
        \frac{1-12\ln2}{96\sqrt{2}\pi}\right)   \om_p^3\,,
\ee
and the second is defined as
\be\label{ETMwg1}   E^{\rm TM}_{\rm wg}=\tilde{E}^{\rm TM}_{\rm wg}
    +\left(-\frac{c}{2\pi^2}+\frac{1}{9\pi^2}+\frac{1-12\ln2}{96\sqrt{2}\pi}\right)   \om_p^3
        +E_{L_1}-E_{L_2}\,.
\ee
The terms, which are subtracted in these two expressions, follow from the definition \Ref{ETM1} of the renormalized vacuum energy. Thereby the definition \Ref{ETMsf1} is   motivated from Eq. \Ref{MEsi2} in the sense, that just the doubled contribution from a plasmon on a single surface is subtracted and
\be\label{MEsf2}     E_{sf}=\tilde{E}_{sf}-2E_{single}
\ee
holds. After that, the subtractions left for \Ref{ETMwg1} are already uniquely determined.
These can be shown to be just that substraction, which makes  $E^{\rm TM}_{\rm wg}$ vanishing for large separation,  $L\to\infty$, thus making  $ E^{\rm TM}_{\rm wg}$ the correctly renormalized contribution to the vacuum energy.   The waveguide contribution can be also split into the TE contribution and a remainder,
\be\label{ETMwg2}   E^{\rm TM}_{\rm wg}=E^{\rm TE}_{\rm wg}+\Delta E_{\rm wg},
\ee
where $E^{\rm TE}_{\rm wg}$ was defined in \Ref{ETEwg2} and $\Delta E_{\rm wg}$ is
\be\label{dEwg2} \Delta E_{\rm wg}=\Delta \tilde{E}_{\rm wg}+
                \left(-\frac{c}{2\pi^2}+\frac{1-12\ln2}{96\sqrt{2}\pi}\right)   \om_p^3
\ee
with $\Delta \tilde{E}_{\rm wg}$ defined in \Ref{dEwg1} taken with $s=0$ and   \Ref{h12} was used.
\begin{figure}[t]
\includegraphics[width=14.5cm]{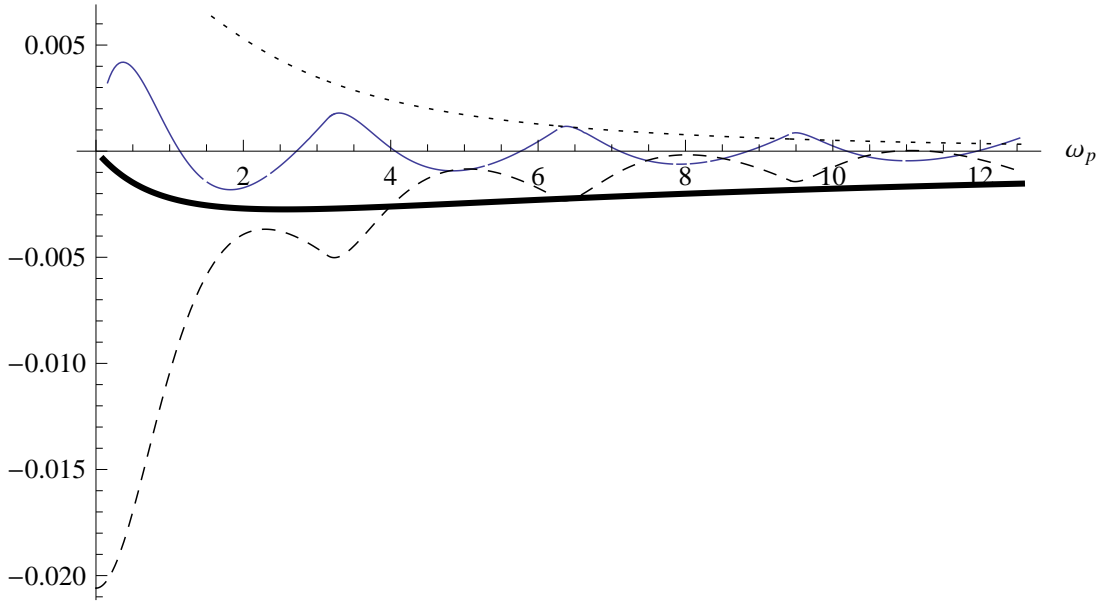}
\caption{The difference between the contributions to the vacuum energies resulting from TM and TE polarizations  from the photonic modes (solid line) and from the waveguide modes (dashed line) as function of $\om_p$ for $L=1$. The  dotted line represents  the sum of the energies of  the surface plasmons. The thick solid line is the difference between the complete contributions from the TM and the TE polarizations.}\label{figdE}
\end{figure}

Collecting from the above formulas, we have a with \Ref{ETM2} and \Ref{ETM3} two representations of the vacuums energy resulting from the TM polarization,
\be\label{ETM4}E^{\rm TM}=E^{\rm TM}_{\rm imag}=E^{\rm TM}_{sf}+E^{\rm TM}_{\rm wg}+E^{\rm TM}_{\rm cont}.
\ee
One is with integration over imaginary frequencies, the other is over real frequencies, i.e., over the real spectrum. Like  the TE case, both can be evaluated numerically and must deliver the same numbers. Indeed, we checked, they do. For technical reasons we represent the results as difference between the TM and  TE cases
\be\label{dE} \Delta E\equiv E^{\rm TM}- E^{\rm TE}.
\ee
Using \Ref{dEimag1}, \Ref{ETMcont2} and \Ref{ETMwg2}, together with \Ref{ETE4}, we get
\be\label{dE1} \Delta E=\Delta E_{\rm imag}=E^{\rm TM}_{sf}+\Delta E_{\rm wg}+\Delta E_{\rm cont}
\ee
The surface modes appear in the TM part only, therefore these remain without TE counterpart which could be subtracted. The constituent energies in Eq. \Ref{dE1} are shown in Fig. \ref{figdE} as functions of the plasma frequency $\om_p$.  Like in the TE case in Fig. \ref{figTE}, there are oscillations and compensation. All curves decrease for large $\om_p$ showing that the difference between TE and TM becomes subleading for large separation or large plasma frequency.

\section{Conclusions}
In this paper we have discussed the role of the electromagnetic modes contributions to the vacuum energy for two half--spaces characterized by a permittivity $\ep$ and separated by a gap of width $L$ with $\ep=1$. In this configuration, the renormalized vacuum energy can be represented as sum of contributions from the TE and the TM polarizations,
\be\label{CE0}E_0=E^{\rm TE}+E^{\rm TM},
\ee
where $E^{\rm TE}$ is given either in terms of integration over imaginary frequency, \Ref{ETE2} for $E^{\rm TE}$ and \Ref{MEimag4} for $E^{\rm TM}$, or in terms  of  real frequencies, \Ref{ETE4} for $E^{\rm TE}$ and \Ref{ETM3} for $E^{\rm TM}$.
Of course, both kinds of representations are equivalent, and for both we derived representations suitable for numerical evaluation. The corresponding numbers have been checked to coincide within the confines of reasonable numerical precision. Results are shown in Figures \ref{figTE} and \ref{figdE}.

The representation in terms of imaginary frequency is, of course, identical to the Lifshitz formula, Eq. \Ref{LF}. As discussed in the Introduction, the representation in terms of real frequencies, considered here for the plasma model, cannot be compared directly with those    representations assuming ${\rm Im}~ \ep(\om)>0$ which can be found in literature.
Restricted to the plasma model considered in this paper,  it is important to stress that the spectrum, i.e., the set of contributing real frequencies, consists of surface (for TM polarization), waveguide and photonic modes. While the   surface modes are well known to contribute, the role of waveguide and photonic modes is less clear in literature. As already mentioned in the Introduction,
in  \cite{ninh70-2-323} only the surface modes were considered, in \cite{gerl71-4-393}, in addition, all evanescent waves  were included.
Also in \cite{bord06-39-6173}, it was not observed that there is a difference between waveguide and photonic modes. However, in \cite{intr07-76-033820} attention was paid to this difference. However, that remained without consequences for the calculations and conclusions of the mentioned papers (and many others) since all calculations were performed in fact on the  imaginary frequency axis and only the contribution from the surface plasmons were calculated on the real frequency axis. The photonic and the waveguide contributions were never calculated separately for the plasma model\footnote{In  \cite{beze07-52-701}  the contributions from the photonic and from the evanescent waves were calculated.}.

In Section 3, with Eqs. \Ref{ETE4} and \Ref{ETM3}, we derived representations of the vacuum energy in terms of real frequencies. The renormalization was done in a way as to ensure decrease for $L\to\infty$. This was archived, in the TE part, by Eq. \Ref{ETE1}, and, for the TM part, by Eq. \Ref{ETM1}. In each case, the subtraction has a term which is independent on the separation and one term which is proportional to $L$. The latter appears as part of the subtraction of the empty space contribution (another piece of this kind was already subtracted during the derivation  of Eq. \Ref{E2}, see  \cite{bord95-28-755} for details). The first subtraction term ensures that the vacuum energy decreases for $L\to\infty$. Also, we mention that this term does not enter the force. In this way, the renormalization done is well justified.

We went the way from the initial mode sum \Ref{E1} to imaginary frequencies, where we performed the renormalization, and back to real frequencies.

Motivated by the initial mode sum \Ref{E1}  in Sections 2, we went the way from real frequencies
 to imaginary frequencies, where we performed the renormalization, and back to real frequencies.
 In   Appendix A we demonstrated how to go the other way round, from imaginary to real frequencies. Here, we started directly from the renormalized vacuum energy \Ref{ETE2} shortening the calculation significantly. It is needless to stress that both ways are equivalent. An interesting by-product is relation \Ref{AAB1} which may serve as a check for the transmission coefficient.

As already mentioned, the two representations, one in terms of imaginary frequencies, the other in terms of real frequencies, coincide and that in terms of imaginary frequencies is just the well known one. In opposite, that in terms if real frequencies, has some unexpected features. First, there are the contributions from the waveguide modes and from the photonic modes already in the TE case. In the TM case, the surface modes come in addition. Second, the waveguide and the photonic modes show oscillations which do not decrease for large separation.
This is clearly seen in Fig. \ref{figTE} for the TE case. For the TM case, the same holds since the difference between both cases, shown in Fig. \ref{figdE}, decreases for $L\to\infty$.

This result is somehow counterintuitive. For large separation (equally, for large $\om_p$), the  half--spaces become ideal conductors and the spectrum consists of the waveguide modes only (take, for example, $\om_p\to\infty$ in Fig. \ref{fig2}). However, for any fixed $L$ (or $\om_p$), we have a sum, $E^{\rm TE}_{\rm wg}+E^{\rm TE}_{\rm cont}$, where the oscillations compensate each other to a large extend. As seen in Eq. \Ref{ETE4}, the sum equals $E^{\rm TE}_{\rm imag}$, Eq. \Ref{ETE4}, which deceases $\sim L^{-3}$.

It must be mentioned, that the attempt to identify the contributions from the physical modes to the vacuum energy works well for small separation. Best it works in the limit  $L\to0$ (non retarded case), where the surface modes of the TM polarization dominate the vacuum energy,
\be\label{CE1}E_0\sim -c_2\frac{\om_p}{L^2}\qquad\mbox{for}\quad L\to0,
\ee
(see Eqs. \Ref{MEimag6}), while the contributions from the waveguide and photonic modes of both polarizations are subleading. However, as shown in \cite{bord06-39-6173}, for increasing separation, there is a compensation between the two surface modes, and further increasing $L$, there is a compensation between the sum of the contributions from the surface modes and from the remaining modes\footnote{The latter compensation was mentioned earlier in \cite{intr05-94-110404}.}. In \cite{bord06-39-6173}, this remaining part was calculated as the difference between the  energy calculated on the imaginary axis and the surface modes's contribution. This remaining part was called 'photonic'. But now we see that it consists of two parts, wageguide and photonic, and we observe also large compensations between these two.

As a problem for further investigations we mention that the compensations between the waveguide and the photonic modes  for large separation, possibly, can be handled analytically. This would make calculation of the vacuum energy in terms of real frequencies easier.

In the present paper we considered the plasma model with permittivity given by Eq. \Ref{ep1}. It should be mentioned  that the separation   into waveguide and photonic modes  will appear also in the easier case of a constant permittivity (one needs either to take $\ep<1$ or to consider a suitable $ep$ in the gap).  And, of course, this separation  will show up if considering more complicated permittivities, at least such without dissipation.

\section*{Acknowledgement}
The author benefited from exchange of ideas by the ESF Research Network
CASIMIR.\\
The author is happy to acknowledge very fruitful discussions with  G. Barton, S. Scheel and S. Buhmann.\\
This work was supported also by the grant   HLP-2011-35 within the {\it Heisenberg-Landau} program
and discussions with I. Pirozhenko, V. Nesterenko, G. Klimchits\-kaya and V. Mostepanenko are acknowledged very much.

\begin{appendix}
\section{From imaginary to real frequencies}
In the main text, in Section 3, we considered the transition from real frequencies in the vacuum energy, to be inserted into the mode sum, Eq. \Ref{E1}, to imaginary frequencies in order to establish the relation with Eq. \Ref{LF}.  In this appendix we consider the reversed transition, from imaginary frequencies as used in  Eq. \Ref{Eimag2}, to real frequencies, as used in Eq. \Ref{ETE3}. Of course, both are equivalent. However, in Section 3 we started from the unrenormalized vacuum energy, whereas here we can start from the already renormalized one, which makes the calculations easier. Further we restrict ourselves to the TE case. The corresponding formulas for the TM case can be written down too, but that would not bring new insights.

\begin{wrapfigure}{r}{0.45\textwidth}
\centering \includegraphics[height=3.4cm]{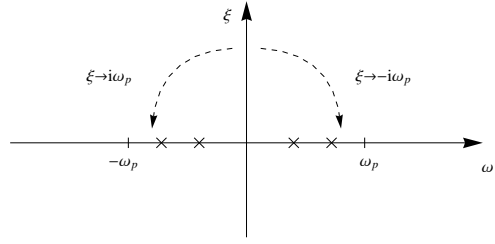}
\caption{The complex $\om$-plane. On the  real axis, the crosses denote the poles corresponding the to waveguide modes. The dashed lines indicate  ways to rotate the integration path.}\label{figA}
\end{wrapfigure}

We start from representation \Ref{Eimag2} of the renormalized vacuum energy in terms of imaginary frequencies,
\be\label{AEimag1}    {E}^{\rm TE}_{\rm imag}=
                   \frac{1}{12\pi^2}\int_0^\infty d\xi\,\xi^3
                    \frac{d}{d\xi}\,\ln t(\xi),
\ee
where we denoted, using \Ref{T5},
\be\label{At1}  t(\xi)\equiv T_1^{\rm TE}(i \xi)=
        \frac{1}{1-\left(\frac{\ka-\xi}{\ka+\xi}\right)^2e^{-2\xi L}}
\ee
with $\ka=\sqrt{\om_p^2+\xi^2}$. In the complex $\om$-plane (see Fig. \Ref{figA}), we integrate in \Ref{AEimag1} over the imaginary axis, $\om=i\xi$. We note that the integrand, considered as function of $\xi$,  has poles in $\xi=\pm i q_j^{\rm TE}$, which correspond to the frequencies \Ref{omj1} of the waveguide modes. Note, that in Eq. \Ref{AEimag1}, the integration over the momenta $\mathbf{k}_{||}$ is already carried out and the relation bet\-ween the momenta and the frequency is given by Eq. \Ref{D2} with formally set ${k}_{||}=0$.

Now we turn the integration path in \Ref{AEimag1} to the right, $\xi\to -i\om$. Using $\xi^3\to  i\om^3$, we get
\bea\label{AEimag2}  &&{E}^{\rm TE}_{\rm imag}\\\nn&&=
                   \frac{i}{12\pi^2}\int_0^\infty d\om\,\om^3
                    \frac{d}{d\om}\,\ln t(-i(\om+i 0)),
\eea
where $'\!\!+i 0'$ indicates that the integration path approaches the real $\om$-axis from above.
We separate this expression into two parts,
\be\label{AEimag3}  {E}^{\rm TE}_{\rm imag}=A+B,
\ee
where
\be\label{AA1} A=\frac{i}{12\pi^2}\int_0^{\om_p} d\om\,\om^3
                    \frac{d}{d\om}\,\ln t(-i(\om+i 0))
\ee
has the integration region $\om\in[0,\om_p]$. This region corresponds to the waves  which are usually called evanescent. Accounting for the poles of the integrand, we represent this integral as the sum of     the residua from passing these poles (entering with a factor $-i\pi$ since $t(-i\om)$ has poles) and a Vp-integral,
\be\label{AA2}  A=\tilde{E}^{\rm TE}_{\rm wg}+
    \frac{i}{12\pi^2}\,\mbox{Vp}\!\!\int_0^{\om_p} d\om\,\om^3  \frac{d}{d\om}\,\ln t(-i\om).
\ee
The residua collect just into the sum over the waveguide modes, Eq. \Ref{Etwg2}, for $s=0$. In the transmission coefficient entering this integral we note that $\ka=\sqrt{\om_p^2-\om^2}$ is real. Hence,
\be\label{At2} t(-i\om)= \frac{1}{1-\left(\frac{\ka+i\om}{\ka-i\om}\right)^2e^{2i\om L}},
\ee
can be rewritten as
\be\label{At3} t(-i\om)=\frac{e^{-i \phi}}{-2i\sin(\phi)}
\ee
with the phase
\be\label{Aphi} \phi=\frac{1}{i}\ln\frac{\ka+i\om}{\ka-i\om}+\om L.
\ee
This allows separating the Vp-integral into real and imaginary parts,
\be\label{AA3}  A=\tilde{E}^{\rm TE}_{\rm wg}+
    \frac{1}{12\pi^2} \int_0^{\om_p} d\om\,\om^3 \frac{d}{d\om}\,\phi
     -\frac{i}{12\pi^2}\,\mbox{Vp}\!\!\int_0^{\om_p} d\om\,\om^3  \frac{d}{d\om}\,
     \ln \left(-2i\sin(\phi)\right).
\ee
The integral containing the phase $\phi$, Eq. \Ref{Aphi}, has no poles in its integrand and it can be calculated easily,
\be\label{AA4} \frac{1}{12\pi^2} \int_0^{\om_p} d\om\,\om^3 \frac{d}{d\om}\,\phi
    =\frac{\om_p^3}{9\pi^2}+\frac{\om_p^4\,L}{48\pi^2}.
\ee
It combines in \Ref{AA3} with $\tilde{E}^{\rm TE}_{\rm wg}$ just into $E^{\rm TE}_{\rm wg}$, Eq. \Ref{ETEwg2}. Hence, the real part of $A$ is simply
\be\label{AA5} {\rm Re}(A)=E^{\rm TE}_{\rm wg}.
\ee
The imaginary part of $A$ is the remaining Vp-integral (the last term in Eq. \Ref{AA3}), which we keep as is for the moment.

Now we turn to the remaining part of the integration region, $\om\in[\om_p,\infty)$, i.e., to part $B$ in \Ref{AEimag3}, which we represent in the form
\be\label{AB1}  B=-\frac{1}{12\pi^2}\int_{\om_p}^\infty d\om\,\om^3
                    \frac{d}{d\om}\,
                \left[ {\rm Im}(\ln t(-i\om))-i\,{\rm Re}(\ln t(-i\om))\right].
\ee
Here we note $\ka=-ik$ with $k=\sqrt{\om^2-\om_p^2}~$ and
\be\label{At4} t(-i\om)=T_1^{\rm TE}(k)
\ee
is just the same transmission coefficient as used in  \Ref{Econt}. In this way, we get for the real part of $B$,
\be\label{AB2} {\rm Re}(B)=-\frac{1}{12\pi^2}\int_{\om_p}^\infty d\om\,\om^3
                    \frac{d}{d\om}\,
                {\rm Im}(\ln T_1^{\rm TE}(k)),
\ee
which differs from Eq. \Ref{Econt} merely by a change of the integration variable. So we have
\be\label{AB3} {\rm Re}(B)=E^{\rm TE}_{\rm cont}.
\ee
Collecting from \Ref{AA5} and \Ref{AB3},  we get from the rotation $\xi\to -i\om$ of the integration path,
\be\label{AETE}     E^{\rm TE}=E^{\rm TE}_{\rm wg}+E^{\rm TE}_{\rm cont},
\ee
which is just the same as \Ref{ETE3}.

The imaginary parts, appearing in $A$ and in $B$, must add up to zero,
\bea\label{AAB} &&\hspace{-1cm}{\rm Im}(A)+{\rm Im}(B)
\nn \\&=&
               \frac{-1}{12\pi^2}\int_0^{\om_p}d\om\,\om^3\frac{d}{d\om}\,\ln\left(-2i\sin(\phi)\right)
                +\frac{1}{12\pi^2}\int_{\om_p}^{\infty}d\om\,\om^3\frac{d}{d\om}\,
                {\rm Re}\left(\ln T_1^{\rm TE}(k)\right),
\nn \\  &=&0,
\eea
since the initial expression, Eq. \Ref{AEimag1}, is real. We mention that the two integrals in \Ref{AAB} can be joined into one using \Ref{At3}, such that the relation
\be\label{AAB1} \frac{1}{12\pi^2}\int_{0}^{\infty}d\om\,\om^3\frac{d}{d\om}\,
                {\rm Re}\left(\ln T_1^{\rm TE}(k)\right)=0
\ee
holds.

There is also the possibility to turn the integration path the other way round, $\xi\to i \om$, i. e., to the left in Fig. \ref{figA}. All calculations done above can be repeated with corresponding changes of signs. As a result, one obtains just the same real parts, i. e., Eq. \Ref{AETE}. The imaginary parts appear also the same as Eq. \Ref{AAB}, but with opposite sign.

In this way, we demonstrated how to go the way back in the complex frequency plane as compared to Section 3. A difference is that we used in Section 3 the variable $k$ for the rotation, in this appendix we use the variable $\om$, which appears to be more instructive. It is needless to stress  that both options are equivalent.  We obtained, as expected, the same relation \Ref{ETE3} as before. The procedure represented in this appendix   appears easier than that in Section 3. This is so since we consider here the   vacuum energy after renormalization, Eq. \Ref{ETE1}. It is interesting to note  that the subtraction terms  in the renormalized energy of the waveguide modes, Eq. \Ref{ETEwg2}, appear also from the integral over the frequencies of the evanescent modes, Eq. \Ref{AA4}, without explicit reference to the renormalization.
\section{Contribution from the surface plasmons at small separation}
In this appendix we calculate the individual contributions from both surface plasmons to the vacuum energy at small separation. These dominate the energy and their sum is the same as given by Eq. \Ref{MEimag6}.

The contribution from a surface plasmon to the vacuum energy is given by Eq. \Ref{MEsf1} with $s=\delta=0$, which we consider now for $\om_p\to0$. For this, we need the momenta $\eta_{sf}$ (see Eq. \Ref{omsf}), which are solutions of Eqs. \Ref{Fc3}. From this equation, it follows that for $\om_p<k_{||}$ the inequality
\be\label{B1} \sqrt{k_{||}^2-\om_p^2},\eta_{sf}<k_{||}
\ee
holds. This motivates, for small $\om_P$ the ansatz
\be\label{B2}   \eta_{sf}=k_{||}-\al_{sf}\om_p^2+\dots\,.
\ee
Inserting the ansatz into Eq. \Ref{Fc3} we get
\bea\label{B3}  \al_s&=&\frac{1}{2k_{||}}\,\frac{1}{1+\coth \frac{k_{||}L}{2}}
\nn\\           \al_a&=&\frac{1}{2k_{||}}\,\frac{1}{1+\tanh \frac{k_{||}L}{2}}
\eea
From here, using \Ref{omsi} and \Ref{omsf}, the expansions of the frequencies follow,
\bea\label{B4} \om_s-\om_{single}&=&
            \left(\frac{1}{\sqrt{1+\coth \frac{k_{||}L}{2}}}-\frac{1}{\sqrt{2}}\right)\om_p+O(\om_p^3),
\nn\\   \om_a-\om_{single}&=&
            \left(\frac{1}{\sqrt{1+\tanh \frac{k_{||}L}{2}}}-\frac{1}{\sqrt{2}}\right)\om_p+O(\om_p^3).
\eea
Inserting these into \Ref{omsi} and \Ref{omsf} and integrating over $\mathbf{k}_{||}$ one comes to
\bea\label{B5}      E_s&=&c_s\frac{\om_p}{L^2}+\dots\,,
\nn\\[5pt]          E_a&=&c_a\frac{\om_p}{L^2}+\dots\,,
\eea
with
\bea\label{B6}  c_s&=&\frac{
        \pi^2-24(1+(-1+\frac12 \ln 2)\ln2)}{24\sqrt{2}\pi}\,,
\nn\\       &\simeq&-0.030576,
\nn\\[8pt]\nn           c_a&=&   \frac{1}{2\sqrt{2}\pi}
            \left[2(-1+\sqrt{2}) +(2-\ln2)\ln2  \right.
\\\nn&&
            \left. +{\rm arcsinh}(1)\left(2(-1+\ln2)-{\rm arcsinh}(1)\right)
            -{\rm Li}_2(3-2\sqrt{2}))
           \right]\,,
            \nn\\&\simeq&0.0266708.
\eea
The sum of these gives Eq. \Ref{MEimag6}.
\end{appendix}

\bibliographystyle{unsrt}
\bibliography{C:/Users/bordag/WORK/Literatur/bib/papers,C:/Users/bordag/WORK/Literatur/Bordag}
\end{document}